\pdfoutput=1
\documentclass[cernpreprint, atlasdraft=false, texlive=2023, UKenglish, texmf, orcidlogo]{atlasdoc}
 
\usepackage[subfigure=true]{atlaspackage}
\usepackage{multicol}
\usepackage{atlasbiblatex}
 
\usepackage{atlasphysics}
 
\graphicspath{{logos/}{figures/}}

\usepackage{ANA-HDBS-2019-10-PAPER-defs}
 
\usepackage{multirow}
\usepackage{bm}

\AtlasTitle{Search for high-mass $W\gamma$ and $Z\gamma$ resonances using hadronic $W/Z$ boson decays from 139 fb$^{-1}$ of $pp$ collisions at $\sqrt{s}=$ 13 TeV with the ATLAS detector}
 
\AtlasAbstract{
A search for high-mass charged and neutral bosons decaying to $W\gamma$  and $Z\gamma$ final states is presented in this paper.
The analysis uses a data sample of $\sqrt{s} = 13$~TeV proton--proton collisions with an integrated luminosity of 139~fb$^{-1}$ collected by the ATLAS detector during LHC Run~2 operation.
The sensitivity of the search is determined using models of the production and decay of spin-1 charged bosons and spin-0/2 neutral bosons.
The range of resonance masses explored extends from 1.0~TeV to 6.8~TeV.
At these high resonance masses, it is beneficial to target the hadronic decays of the $W$ and $Z$ bosons because of their large branching fractions. The decay products of the high-momentum $W/Z$ bosons are strongly collimated and boosted-boson tagging techniques are employed to improve the sensitivity.
No evidence of a signal above the Standard Model backgrounds is observed, and upper limits on the production cross-sections of these bosons times their branching fractions to $W\gamma$ and $Z\gamma$ are derived for various boson production models.
}
 
\author{ATLAS Collaboration}
 
\AtlasRefCode{HDBS-2019-10}
 
\PreprintIdNumber{CERN-EP-2022-241}
 
\AtlasJournalRef{JHEP 07 (2023) 125}
\AtlasDOI{10.1007/JHEP07(2023)125}

\journal{JHEP}

\begin{document}

\maketitle

\section{Introduction}

Speculations about physics phenomena beyond those described by the Standard Model (SM) often result in the introduction of new bosons, due to either additional gauge symmetries or postulated extensions of the Higgs sector~\cite{TheoryTC,LittleHiggs,Pappadopulo:2014qza}.
The high-energy proton--proton ($pp$) collisions provided by the Large Hadron Collider (LHC) make it possible to produce these new bosons with masses up to approximately one hundred times the mass of the SM $W$ and $Z$ bosons.
A broad range of beyond-the-SM (BSM) scenarios can therefore be tested with experiments at the LHC that search for high-mass charged and neutral bosons.
 
Some of the BSM theories predict new charged $X^{\pm}$ and neutral $X^{0}$ bosons~\cite{Pappadopulo:2014qza,Artoisenet:2013puc}.
From an experimental perspective, $W\gamma$ or $Z\gamma$ final states are attractive since a high-energy photon signature efficiently selects signal events and rejects background.
For bosons with masses of the order of \TeV, decays of the type $X^{\pm}\to$ $W^{\pm}\gamma$  or $X^0\to$ $Z\gamma$ result in a highly boosted $W$ or $Z$ boson, where the decay products of such a boson are very collimated.
This analysis targets the hadronic decay modes of $W$ and $Z$ bosons to quark--antiquark pairs reconstructed as large-radius (large-$R$) jets that have a two-prong structure identified using jet-substructure information~\cite{JETM-2018-03}.
The complete reconstruction of the  $W\gamma$  or $Z\gamma$  final state can then be used to determine the mass and other properties of the new bosons.
 
This paper presents searches for massive $X^{\pm}$ and $X^0$ bosons using 139~fb$^{-1}$ of $pp$ collisions at a centre-of-mass energy ($\sqrt{s}$) of 13~\TeV\ recorded with the ATLAS detector.
The searches assume that the decay width of the heavy bosons is small compared to the experimental resolution, but are otherwise generic, looking for any excess of events above smooth SM background $W\gamma$ and $Z\gamma$ invariant mass spectra.
The measurements are compared with the predictions of models of the production and decay of spin-1 charged bosons and spin-0/2 neutral bosons.
These include $q \bar{q'}$ annihilation production of spin-1 $X^{\pm}\to$ $W^{\pm}\gamma$, gluon--gluon fusion production of spin-0 $X^0\to$ $Z\gamma$,
and both gluon--gluon fusion and $q \bar{q}$ annihilation production of spin-2 $X^0\to$ $Z\gamma$.
A boson mass ($m_{X}$) range from 1.0 to 6.8~\TeV\ is covered by these searches.
 
Previous searches for bosons of mass greater than 1.0~\TeV\ decaying to  $W\gamma$ and  $Z\gamma$ final states have been carried out at the LHC by the ATLAS~\cite{EXOT-2013-09,EXOT-2016-02,EXOT-2016-30} and CMS~\cite{CMS-EXO-16-021,CMS-EXO-16-025,CMS-EXO-17-005,CMS-EXO-20-001} Collaborations.
Compared to the previous ATLAS search based on 36.1~fb$^{-1}$ of Run~2 $\sqrt{s} = 13$~\TeV\ $pp$ collision data~\cite{EXOT-2016-30}, the search reported
in this paper achieves better sensitivity in part by including the entire dataset collected by the ATLAS experiment during Run~2.
In addition to the four times larger dataset, the search is further improved by a factor of two via optimising the identification of the hadronic decays of highly boosted $W$ and $Z$ bosons.


\section{ATLAS detector}

The ATLAS experiment is a multipurpose detector~\cite{PERF-2007-01} having a forward-backward symmetric cylindrical geometry and almost 4$\pi$ coverage in solid angle.
The inner tracking detectors are immersed in a 2~T magnetic field produced by a thin superconducting solenoid.
The tracking detectors cover a pseudorapidity\footnote{ATLAS uses a right-handed coordinate system with its origin at the nominal interaction point (IP) in the centre of the detector and the $z$-axis along the beam pipe.
The $x$-axis points from the IP to the centre of the LHC ring, and the $y$-axis points upward. Cylindrical coordinates $(r,\phi)$ are used in the transverse plane, $\phi$ being the azimuthal angle around the $z$-axis.
The pseudorapidity is defined in terms of the polar angle $\theta$ as $\eta=-\ln\tan(\theta/2)$.} range $|\eta|<2.5$
using a combination of silicon pixel detectors closest to the beam pipe, followed by silicon microstrip trackers and an outer transition radiation tracker.
The innermost layer, known as the insertable B-layer~\cite{ATLAS-TDR-19,PIX-2018-001}, provides high-resolution hits at small radius to improve the tracking performance.
 
The inner tracking detectors are surrounded  by calorimeters and a muon spectrometer.
The electromagnetic (EM) calorimeter is a lead/liquid-argon (LAr) sampling calorimeter with high granularity. Its barrel~($|\eta|<1.475$) and endcap~($1.375<|\eta|<3.2$) components provide EM energy measurements of electrons and photons up to a pseudorapidity $|\eta|=3.2$.
In the range used for precision measurements of electrons and photons ($|\eta|<2.5$ excluding a transition region $1.37<|\eta|<1.52$),
the EM calorimeter is segmented into three layers along the shower depth, providing excellent measurements of photon properties and allowing precise photon identification.
A steel/scintillator-tile hadronic calorimeter covers the central pseudorapidity range $|\eta|<1.7$.
The endcap and forward regions are instrumented up to $|\eta|=4.9$ with LAr calorimeters for EM and hadronic energy measurements.
 
The muon spectrometer (MS) comprises separate trigger and
high-precision tracking chambers measuring the deflection of muons in a magnetic field generated by superconducting air-core toroidal magnets.
The field integral of the toroids ranges between \num{2.0} and \SI{6.0}{\tesla\metre}
across most of the detector.
A set of precision chambers covers the region \(|\eta| < 2.7\) with three layers of monitored drift tubes,
complemented by cathode-strip chambers in the forward region, where the background is highest.
The muon trigger system covers the range \(|\eta| < 2.4\) with resistive-plate chambers in the barrel, and thin-gap chambers in the endcap regions.
 
Events are selected from the LHC's $pp$ bunch crossings, which occur at a rate of 40~MHz, by a first-level trigger implemented in custom hardware followed by a software-based high-level trigger that employs algorithms similar to those used in offline event reconstruction~\cite{TRIG-2016-01}.
The first-level trigger selects events at a rate of 100~kHz by using a subset of detector information, with the high-level trigger then accepting events for offline analysis at the rate of about 1~kHz.
An extensive software suite~\cite{ATL-SOFT-PUB-2021-001} is used in data simulation, in the reconstruction and analysis of real and simulated data, in detector operation, and in the trigger and data acquisition systems of the experiment.


\section{Data collection and Monte Carlo event simulation}

\subsection{Data samples}
The data used for this analysis were collected by the ATLAS detector from 2015 to 2018 when the LHC provided $pp$ collisions at $\sqrt{s}=13$~\TeV.
Events were selected using a single-photon trigger with loose photon identification requirements based upon EM calorimeter cluster shower-shape variables~\cite{TRIG-2018-05}.
The trigger with a photon transverse energy ($E_{\mathrm{T}}^{\gamma}$) threshold of 140~\GeV\ is fully efficient for events used in this search.
In addition to the trigger selection, events are required to have at least one offline reconstructed signal photon matched to the object that fired the photon trigger.
After requiring that all detector systems were recording high-quality data, the final dataset has an integrated luminosity of 139~fb$^{-1}$~\cite{DAPR-2018-01,ATLAS-CONF-2019-021}.


\subsection{Monte Carlo simulation}
\label{sec:mc}
Monte Carlo (MC) event generators were used to simulate SM background events and BSM heavy-boson signal events.
These simulated event samples are used to optimize the event selection for the new-BSM-boson search and validate the parameterization of the templates used to fit the $W/Z+\gamma$ mass distributions.
The largest background is due to single-photon production in association with jets ($\gamma$+jets) where the jet fulfils the boson-tagging criteria used to identity the large-$R$ jets from $W/Z$ boson hadronic decays.
These events were simulated using the \SHERPA[2.2.2] generator~\cite{Sherpa:2019gpd}, with up to two additional parton emissions included at next-to-leading-order (NLO) precision and up to four additional partons at leading-order (LO) precision.
The matrix elements of these events were calculated with the Comix~\cite{Comix} and \OPENLOOPS~\cite{OpenLoop1,OpenLoop2} libraries and then matched to the \SHERPA parton shower~\cite{SherpaShower} using the MEPS@NLO prescription~\cite{MEPS3,MEPS4,MEPS1,MEPS2}.
The \NNPDF[3.0nnlo]~\cite{NNPDF30} parton distribution function (PDF) set was used to describe the parton distributions in the incoming protons.
The irreducible SM background from the hadronic decays of $W$ and $Z$ bosons produced with a radiated photon was simulated at LO precision with the \SHERPA[2.1.1] generator, and the parton distributions were modelled with the CT10 PDF set~\cite{CT10}.
The SM $t\bar{t}$+$\gamma$ process was simulated with a matrix element at LO with \MGNLO[2.3.3]~\cite{Madgraph}, followed by \PYTHIA[8.186]~\cite{pythia8.1} for the parton showering.
The \NNPDF[2.3lo] PDF set~\cite{NNPDF23} and a set of tuned parameters called the A14 tune~\cite{ATL-PHYS-PUB-2014-021} were used for this
$t\bar{t}$+$\gamma$ event generation.

Various samples of simulated BSM boson signal events are used to optimize
the event selection criteria and to estimate the acceptance and efficiency for the detection of the \XtoWg{} and \XtoZg{} signals.
The production of the $X^{\pm}$ and $X^0$ bosons was modelled in a narrow-width approximation where the natural width of the bosons is much smaller than the
expected experimental resolution of the invariant mass of the $W^{\pm}\gamma$ and $Z\gamma$ resonances.
 
The production of a spin-0 boson decaying into $Z\gamma$ was simulated in gluon--gluon fusion, \ggXZg~\cite{Powheg_ggH}.
This process was modelled with the MC generator \POWHEGBOX[v2]~\cite{POWHEG} at NLO precision as used for SM $H\to Z\gamma$ production, with the Higgs boson mass varied.
The CT10 PDF set was used to generate these events.
The parton showering was modelled with \PYTHIA[8.212]~\cite{Pythia8.2} with the AZNLO tune~\cite{STDM-2012-23}.
 
The spin-1 resonance \qqXWg{} signal process event generation utilized the heavy-vector-triplet framework~\cite{Pappadopulo:2014qza} for event kinematic modelling.
The simulations of the spin-2 \ggXZg{} and \qqXZg{} signal events are based on a resonance model benchmarked from the Higgs characterization model framework with s-channel direct couplings between the spin-2 heavy resonance and the SM $Z$ boson and the $\gamma$~\cite{Allanach:2007ea, Falkowski:2016glr, Dillon:2016fgw}.
The \MGNLO[v2.3.3] MC generator was used at LO precision, followed by \PYTHIA[8.212] for the parton showering with the \NNPDF[2.3lo] PDF set and the A14 tune.
In these models the $W$ ($Z$) boson is produced longitudinally (transversely) polarized.
In samples with \PYTHIA used for parton showering, decays of $c$- and $b$-hadrons were simulated with \EVTGEN[1.2.0]~\cite{EvtGen}.
 
The resulting MC event samples were processed using a detailed simulation of the ATLAS detector with \GEANT~\cite{GEANT4,ATLASFullSim}, and then passed through the same reconstruction algorithms as those used for the data. 
Effects of multiple $pp$ collisions (pile-up) are included during reconstruction by overlaying inelastic events simulated with \PYTHIA[8.186] using the A3 tune~\cite{ATL-PHYS-PUB-2016-017} and the \NNPDF[2.3lo] PDF set.
These minimum-bias events are overlaid with multiplicity distributions that approximately match the pile-up observed in the data.
A pile-up reweighting approach is then performed to correct for the residual difference between simulation and data in the analysis.

A summary of the MC generators used for the SM and BSM processes is given in Table~\ref{tab:MC}.

\begin{table}[hbt!]
\scriptsize
\begin{center}
\caption{Generators used for the simulation of SM backgrounds and BSM signals.}
\label{tab:MC}
\begin{tabular}{l|c|c|c|l}
\hline
\hline
\textbf{Process} & \textbf{Matrix element generator} & \textbf{QCD order} & \textbf{PDF}& \textbf{Parton shower}\\
\hline
\multicolumn{5}{c}{SM backgrounds}  \\
\hline
SM $\gamma$+jets&\SHERPA[2.2.2] & NLO & \NNPDF[3.0nnlo] & \SHERPA MEPS@NLO \\
SM $W\gamma$ and $Z\gamma$ & \SHERPA[2.1.1] & LO & CT10 & \SHERPA MEPS@LO \\
SM $t\bar{t}$+$\gamma$     & \MGNLO[2.3.3]   & LO & \NNPDF[2.3lo] & \PYTHIA[8.186] + \EVTGEN[1.2.0] \\
\hline
\hline
\multicolumn{5}{c}{Signals} \\
\hline
Spin-0 $gg \to X^0 \to Z\gamma$ & \POWHEGBOX[v2] & NLO & CT10 & \PYTHIA[8.212] + \EVTGEN[1.2.0] \\
Spin-2 $gg \to X^0 \to Z\gamma$       & \MGNLO[2.3.3] & LO& \NNPDF[2.3lo] & \PYTHIA[8.212] + \EVTGEN[1.2.0] \\
Spin-2 $q\bar{q} \to X^0 \to Z\gamma$ & \MGNLO[2.3.3] & LO & \NNPDF[2.3lo] & \PYTHIA[8.212] + \EVTGEN[1.2.0] \\
Spin-1 $q\bar{q'} \to X^{\pm} \to W^{\pm}\gamma$ & \MGNLO[2.3.3] & LO & \NNPDF[2.3lo] & \PYTHIA[8.212] + \EVTGEN[1.2.0] \\
\hline
\hline
\end{tabular}
\end{center}
\end{table}


\section{Event reconstruction}

Events are required to pass a loose identification photon trigger with a transverse energy ($E_{\mathrm{T}}^\gamma$) threshold of 140~\GeV.
Each of these events is then processed through offline particle reconstruction to identify high-\ET
photons and to search for jets that pass a $W/Z$ boson tagging requirement.
The details of the photon, jet  and $W/Z$ boson reconstruction and identification are
described in this section, along with the categorization applied to define the signal regions for the \XtoWg{} and \XtoZg{} BSM boson searches.

\subsection{Particle reconstruction}
 
Photon candidates are reconstructed from clusters of energy in the EM calorimeter and classified either as converted photons (those with a reconstructed vertex
consistent with a $\gamma \to e^+e^-$ conversion) or as unconverted photons~\cite{EGAM-2018-01}.
The photon identification algorithm uses shower shape variables measured from both the fine segmentation of the inner layers of the EM calorimeter and the
outer layers of the EM and hadronic calorimeters to suppress background from photons from neutral meson decays in jets.
For this analysis, \textit{tight} photons are selected, with a measured photon identification efficiency greater than 90$\%$ (95$\%$)
for unconverted (converted) photon candidates with $E_{\mathrm{T}}^\gamma > 200\,\GeV$~\cite{EGAM-2018-01}.
 
To further reduce backgrounds from jets, an isolation requirement~\cite{EGAM-2018-01} is imposed on the photons, using
the transverse energy ($E_{\mathrm{T}}^{\mathrm{iso}}$) deposited in the EM calorimeter within a cone of size $\Delta R \equiv \sqrt{(\Delta\eta)^{2}+(\Delta\phi)^{2}} = 0.4$ centred on the photon candidate,
excluding the photon transverse energy within an area  $\Delta\eta \times \Delta\phi = 0.125 \times 0.175$.
After corrections for photon energy leakage into the isolation cone and contributions from the underlying event and pile-up interactions, the photon
isolation transverse energy $E_{\mathrm{T}}^{\mathrm{iso} }$ is required to be less than $0.022 \times E_{\mathrm{T}}^{\gamma} + 2.45\,\GeV$.
For the signal photons passing the reconstruction and identification requirements, the isolation efficiency is approximately 98$\%$.
Events selected for analysis must have at least one isolated  photon candidate with $E_{\mathrm{T}}^\gamma> 200~\GeV$ and $|\eta^\gamma|< 1.37$.
The $\eta$ requirement is motivated by the fact that the photon from a signal event tends to be more central than those from the background.

Jets are reconstructed using charged-particle tracks and calorimeter energy clusters, combining their information to optimize the measurement of the jet direction and energy~\cite{JETM-2018-06}.
The clustering method is that of the anti-\kt{} algorithm~\cite{antiKt,FastJet} with radius parameter $R = 1.0$.
In order to reduce contributions to the jet from pile-up, a trimming algorithm~\cite{Krohn:2009th} is applied, which removes contributions from sub-jets clustered using the \kt{} algorithm~\cite{kT} with $R = 0.2$ if they carry less than 5$\%$ of the jet's transverse momentum.
The jets are calibrated to the level of stable final-state particles using MC simulations~\cite{JETM-2018-02}.
Jets are selected if they have a transverse momentum $p_{\mathrm{T}}^J> 200\,\GeV$ and are within a pseudorapidity region $|\eta^{J}|< 2.0$, where the inner tracker has good charged-particle tracking coverage.
The jets are also required to be separated from photons by $\Delta R(J,\gamma) > 1.0$.
 
A $W$ or $Z$ boson produced by the decay of a boson with a mass of the order of a few \TeV\ is highly boosted, with the di-quark decay products often forming a single large-$R$ jet.
The characteristics of these di-quark jets can be used to distinguish $W/Z$ bosons from a large background of jets originating from single quarks or gluons.
The main distinguishing features are the jet mass and the presence of two-prong substructure within the jet.
 
\begin{figure}[htbp]\centering
\begin{center}
\subfigure[]{\includegraphics[width=0.44\textwidth]{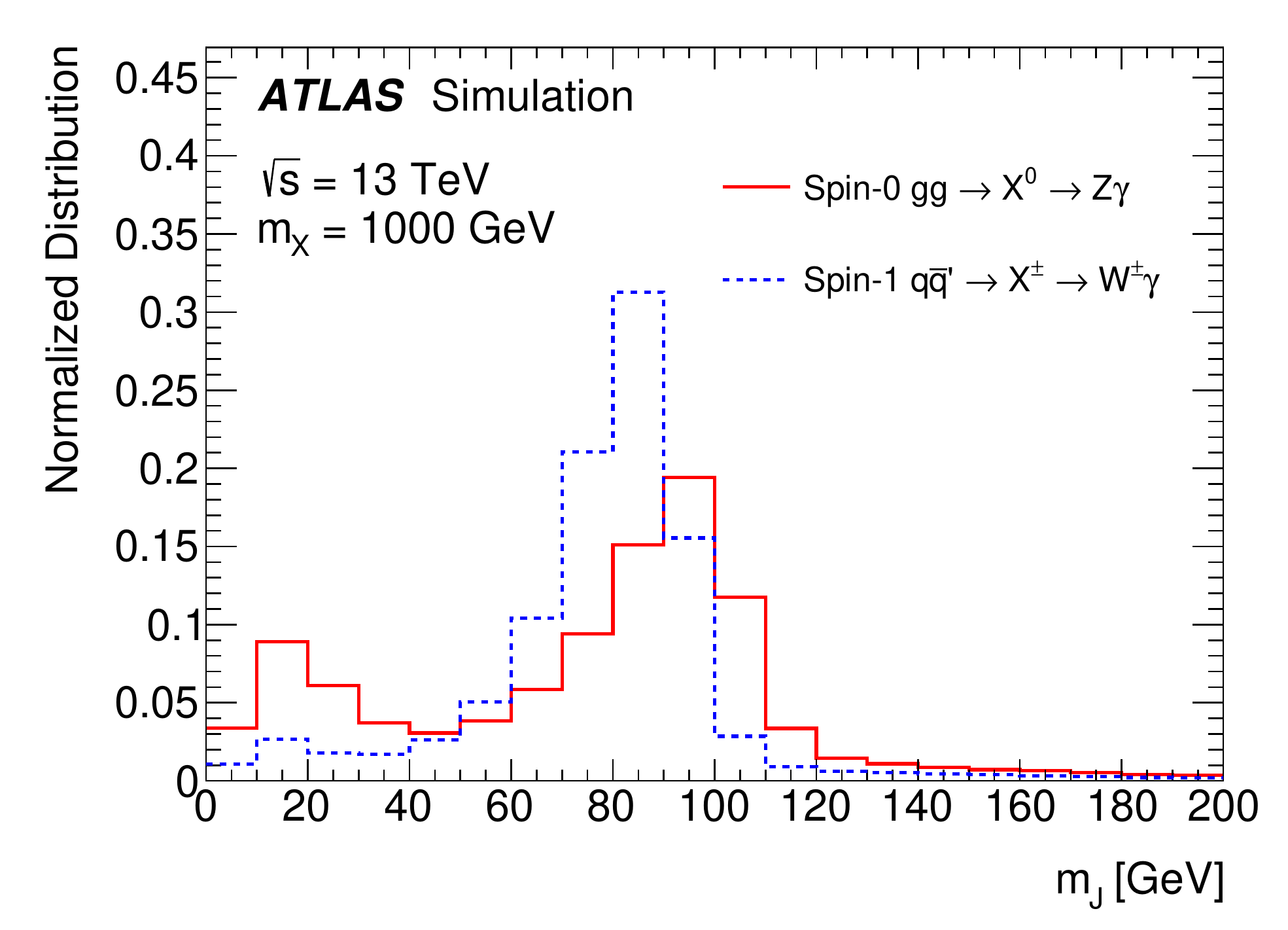} }
\subfigure[]{\includegraphics[width=0.44\textwidth]{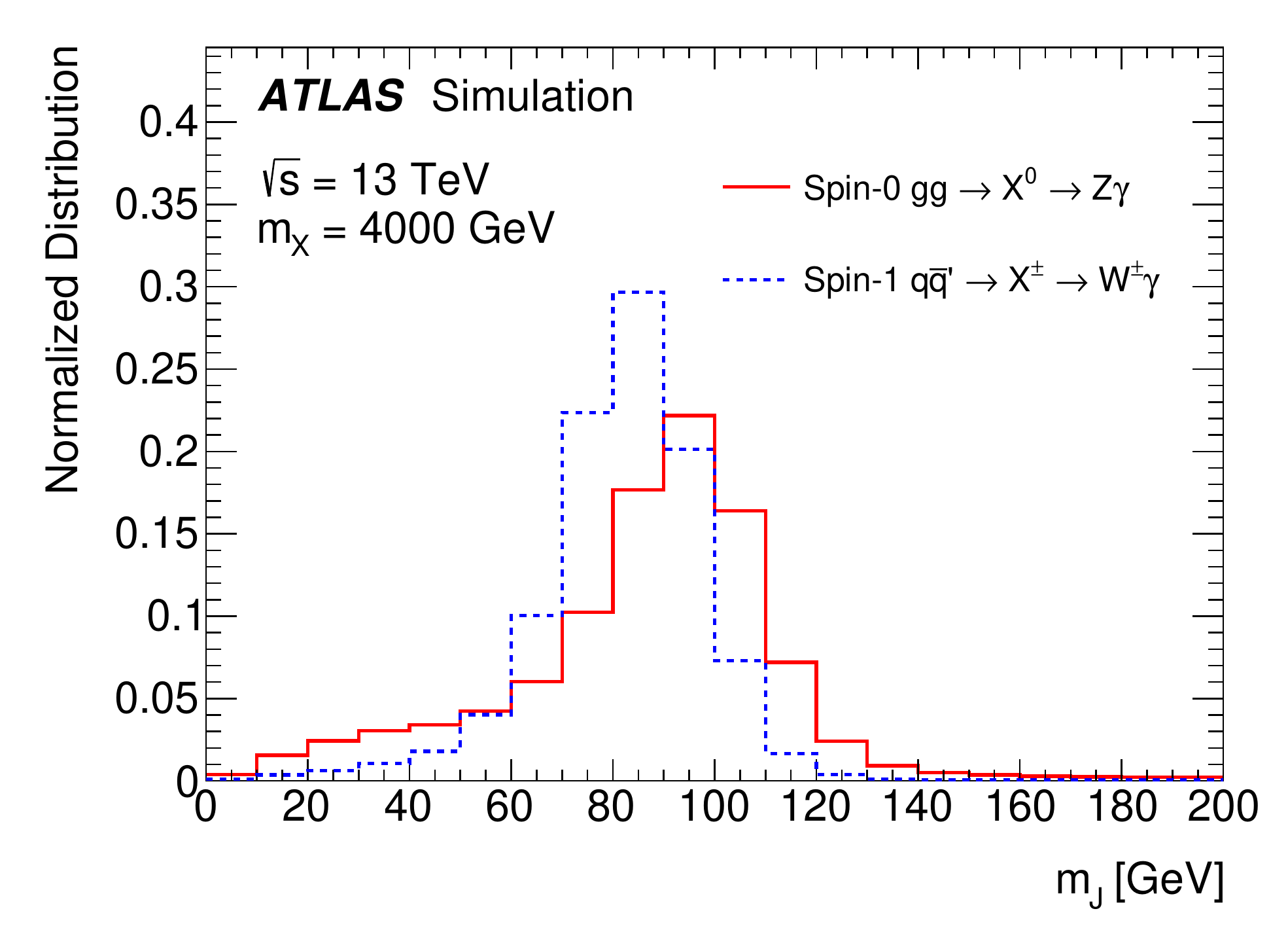} }
\caption{The jet mass distribution of large-$R$ jets originating from the hadronic decay
of $W$ and $Z$ bosons produced from the decay of BSM bosons with mass
(a) $m_X = 1000\, \GeV$ and (b) $m_X = 4000\,\GeV$. The decays simulated are for
the production models \qqXWg{} with a spin-1 resonance $X^{\pm}$
and \ggXZg{} with a spin-0 resonance $X^{0}$.
The $Z$ bosons from $Z\gamma$ decays of spin-2 resonances have jet mass distributions very similar to those shown for spin-0 resonances.
}
\label{fig:SigJetMass}
\end{center}
\end{figure}

The jet mass ($m_{J}$) is calculated using a combination of particle four-momenta
measured from charged-particle tracks and calorimeter cell energies~\cite{ATL-PHYS-PUB-2017-015}.
The jet mass resolution ranges from 8$\%$ to 15$\%$ for jets with transverse momentum between 500 and 2500~\GeV, respectively.
Reconstructed jet mass distributions from simulated hadronic decays of $W$ and $Z$ bosons are shown in Figure~\ref{fig:SigJetMass}.
The low mass tail is caused by events where the decay products from a $W$ or $Z$ boson are not fully captured in the large-$R$ jet.
The effects are different for $Z$ and $W$ bosons since the $Z$ boson from $X^{0}$ decay is transversely polarized whereas the $W$ boson from $X^{\pm}$ decay is longitudinally polarized.
The jet mass is required to be in a window around the boson mass where the window's size is optimized as a function of $p_{\mathrm{T}}^J$ to maximize the significance of the $W$ or $Z$ boson selection over multijet backgrounds~\cite{HDBS-2018-10}.
The size of the mass window increases from about 20 to 50~\GeV\ as \pTJ{} increases from 500 to 2500~\GeV.
For a large-$R$ jet with $\pT<500\,\GeV$ ($\pT>2500\,\GeV$), the criterion defined at $\pT=500\,\GeV$ ($\pT=2500\,\GeV$) is applied.
 
The two-prong jet substructure from hadronic $W/Z$ boson decays is identified using the energies and pairwise angular distances between clusters of particles within the large-$R$ jets.
This is quantified with a variable $D_2 $ defined as the ratio $\epsilon_3/[\epsilon_2]^3$ of $N$-point energy correlation functions $\epsilon_N$ computed from the jet constituents~\cite{D2Ref1,D2Ref2}.
This variable exploits the sensitivity of $\epsilon_2$ to the hadronic shower produced from a single quark or gluon versus $\epsilon_3$, which is sensitive to the two hadronic-jet clusters produced from the di-quark decay of $W/Z$ bosons.
Studies using simulations and data were used to choose the requirements on $D_2$ that optimize the $W/Z$ boson identification significance~\cite{HDBS-2018-10}.
The chosen upper limit on $D_{2}$ varies from 1.0 at low jet \pt{} to slightly above 2.0 at high jet \pt{} for the $W/Z$ hadronic jets used in this analysis.
 
For the fraction of $Z$ bosons that decay into $b\bar{b}$, the purity of the selection can be further improved by applying $b$-hadron identification requirements.
A tagging algorithm is used that exploits the long lifetime of $b$-hadrons, which leads to tracks with large impact parameters and to secondary vertices.
The outputs of three $b$-tagging techniques are combined into a single multivariate discriminant, called MV2c10, allowing the selection of $b$-hadrons with various efficiencies and background rejections~\cite{PERF-2016-05}.
This $b$-tagging algorithm is applied to variable-radius (VR) track-jets associated with the large-$R$ jet, as determined by the ghost-association algorithm~\cite{Cacciari:2007fd}.
The VR track-jets are reconstructed from ID tracks using the anti-\kt{} algorithm with a variable radius parameter $R$ that ranges between 0.02 and 0.4 depending on the jet \pt \cite{ATL-PHYS-PUB-2017-010}.
The tagging efficiency is determined with simulated $t\bar{t}$ events and corrected to the measurement in data~\cite{FTAG-2018-01}.
A working point with a $b$-tagging efficiency of 70$\%$ is used.
Two VR track-jets are required to pass this $b$-tagging requirement to select $Z\to b\bar{b}$ events.

\subsection{Event selection and categorization}
The events selected are required to have a photon with $\ETg> 200\,\GeV$ and $|\eta^\gamma|< 1.37$ and a jet with $\pTJ > 200\,\GeV$ and $|\eta^J|< 2.0$, using the identification
criteria described above. These selection criteria are called the \enquote{baseline selection} for this analysis.
The $pp$ interaction vertex selected for reconstruction of these physics objects is the one with the highest sum of the $\pT^2$ of the tracks coming from the vertex.
If multiple photons or jets satisfy the photon/jet selection criteria, those with the highest transverse energy or momentum are used.
The search considers resonances with masses larger than 1~\TeV.
Below this mass, the signal selection efficiency drops significantly because of the criteria used to select the hadronic decays of the $W/Z$ bosons, and searches for $pp\to X\to W/Z+\gamma$ with leptonic $W/Z$ boson decays are more sensitive.
The search range is limited to 6.8~\TeV\ using the highest-mass $\gamma$+jet event observed in data.
The selected events are further sorted into exclusive categories of different $W$ and $Z$ boson identification purities to maximize the signal sensitivity.
 
\begin{figure}[tbph]
\begin{center}
\includegraphics[width=\textwidth]{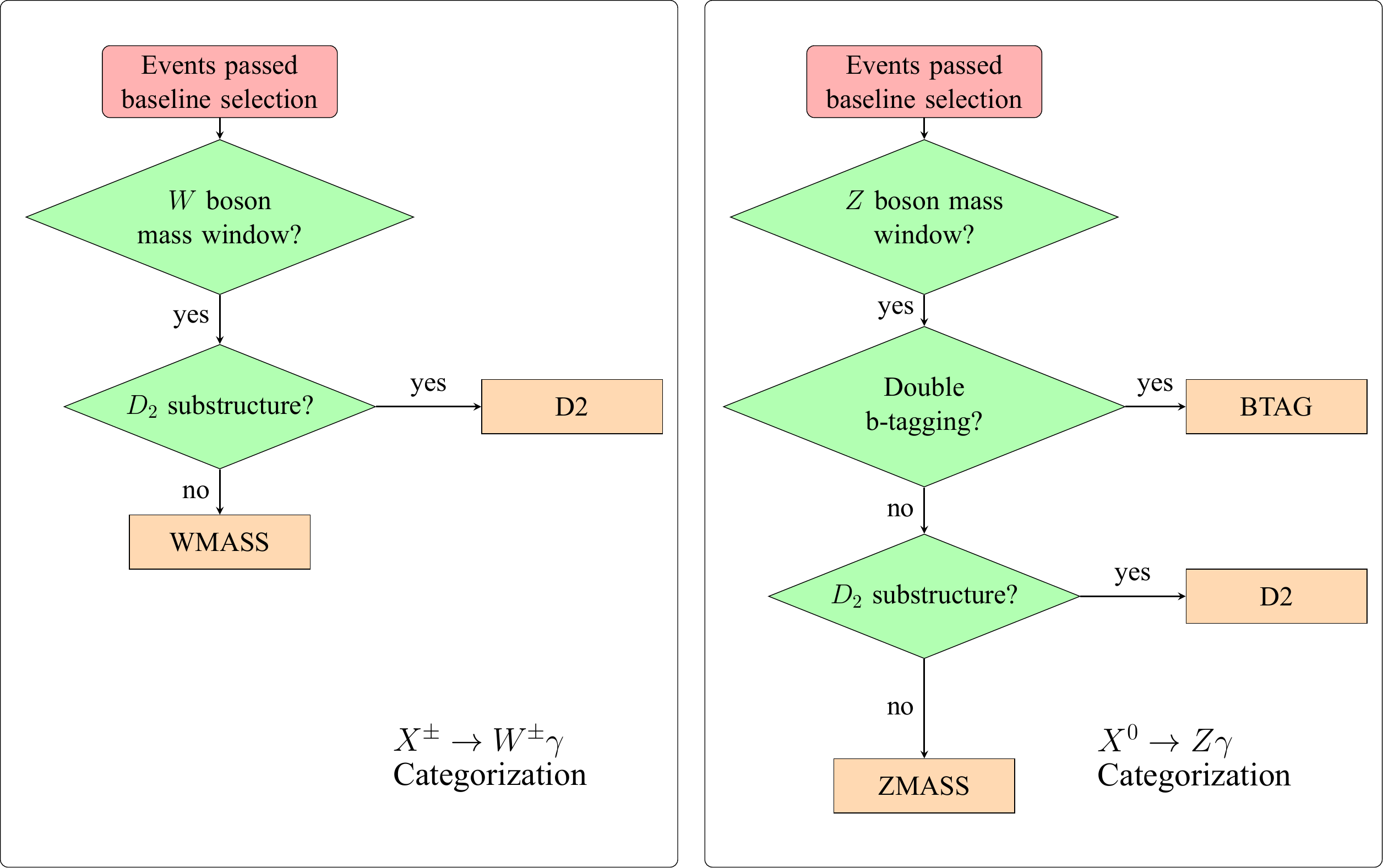}
\end{center}
\caption{The flow charts of event categorization of
\XtoWg{} and
\XtoZg{}.
}
\label{fig:flowchart}
\end{figure}

For the \XtoWg{} search, two categories are defined according to the $D_{2}$ and jet mass criteria shown below, with the category designation indicated in parentheses.
\begin{itemize}
\item{\textit{pass} $D_{2}$ and $W$ boson mass selection (D2)},
\item{\textit{fail} $D_{2}$ and \textit{pass} $W$ boson mass selection (WMASS)}.
\end{itemize}
For the \XtoZg{} search, three categories are defined, based on the $b$-tagging, $D_{2}$ and jet mass criteria shown below.
\begin{itemize}
\item{\textit{pass} two $b$-tagged sub-jets and \textit{pass} $Z$ boson mass selection (BTAG)},
\item{\textit{fail} two $b$-tagged sub-jets: \textit{pass} $D_{2}$ and $Z$ boson mass selection} (D2),
\item{\textit{fail} two $b$-tagged sub-jets: \textit{fail} $D_{2}$ and \textit{pass} $Z$ boson mass selection (ZMASS)}.
\end{itemize}
Figure~\ref{fig:flowchart} illustrates the categorization of \XtoWg{} and \XtoZg{} events.
 
\begin{figure}[htbp]\centering
\begin{center}
\subfigure[]{\includegraphics[width=0.44\textwidth]{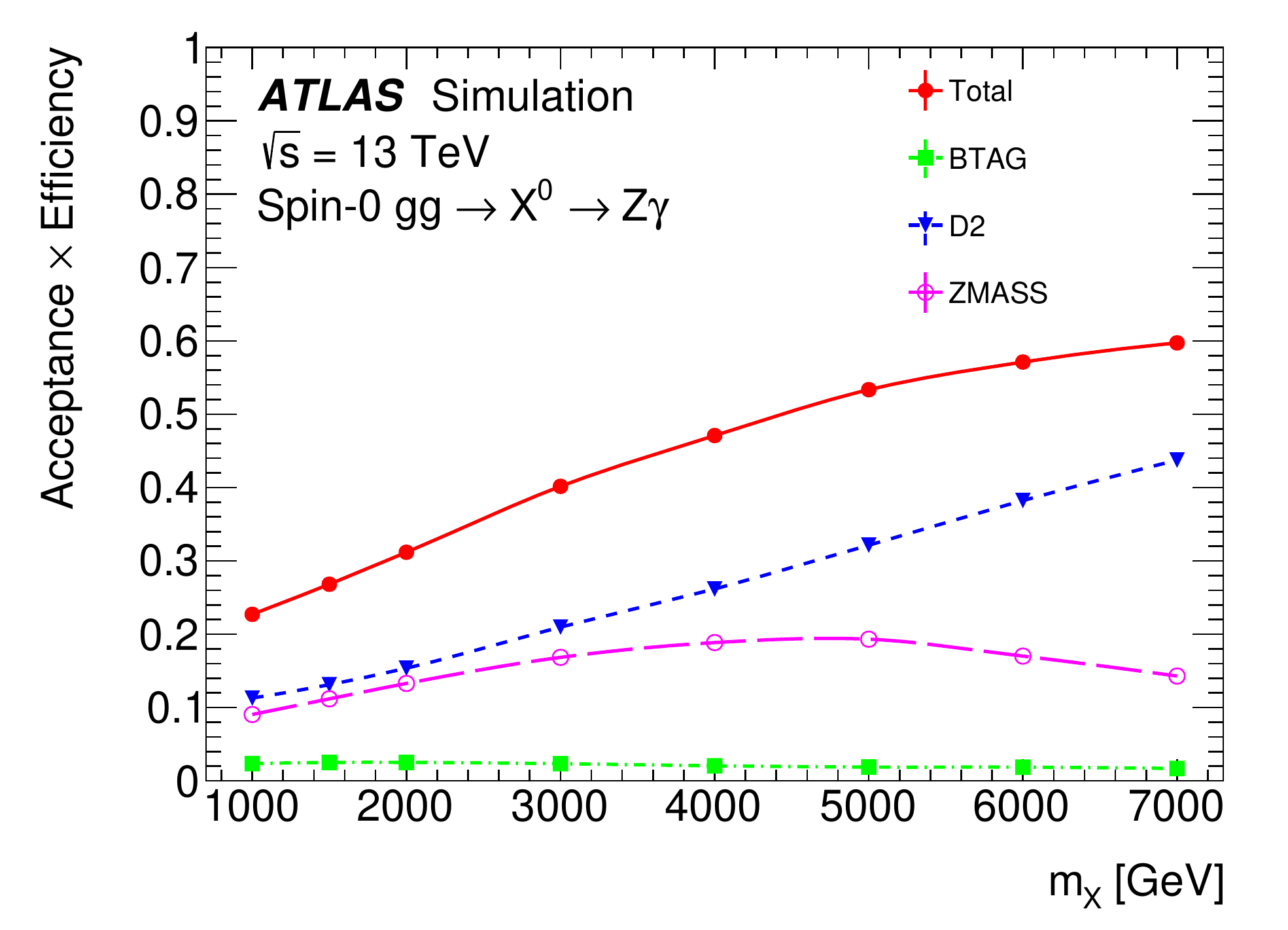} }
\subfigure[]{\includegraphics[width=0.44\textwidth]{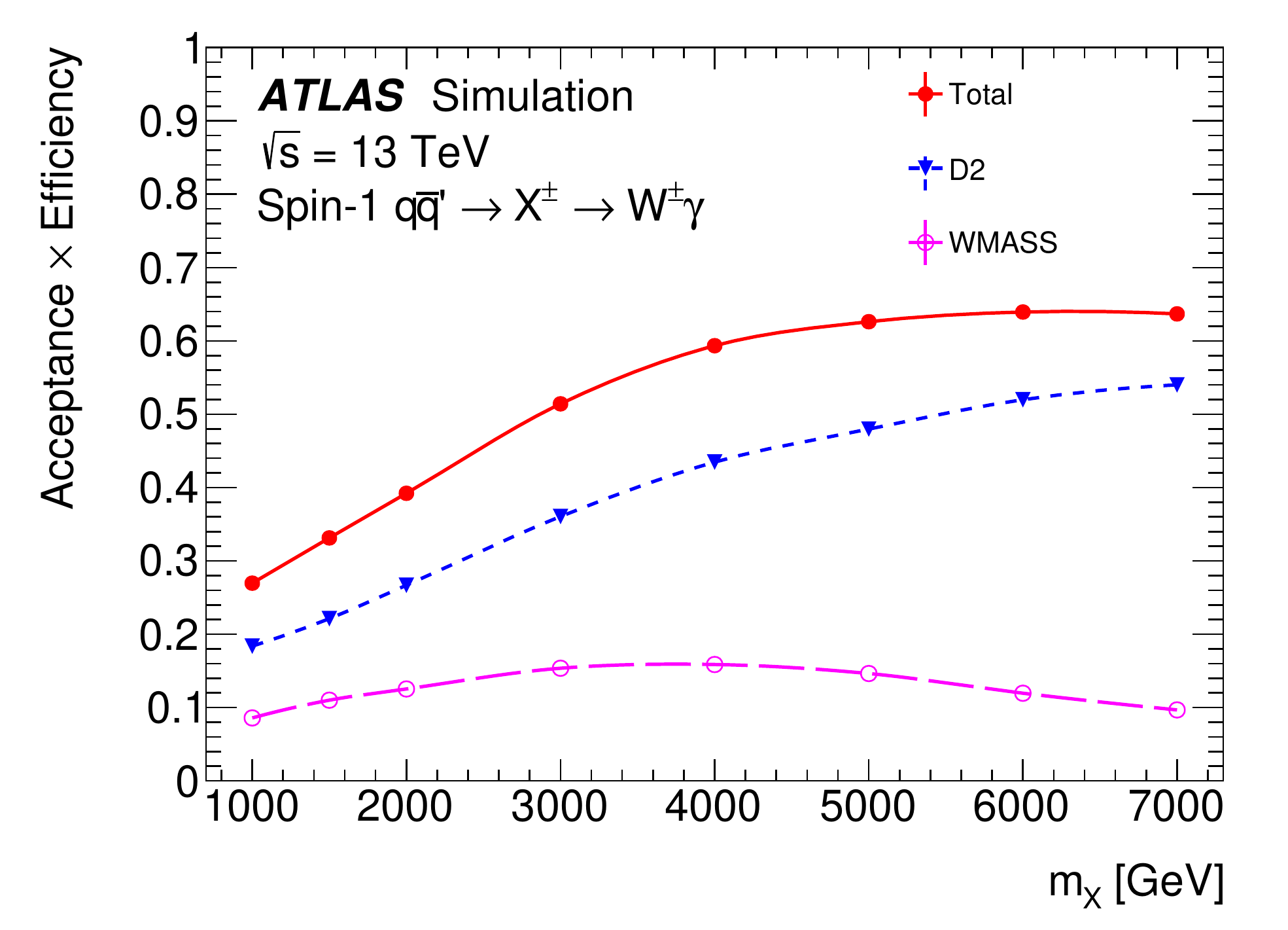} }
\subfigure[]{\includegraphics[width=0.44\textwidth]{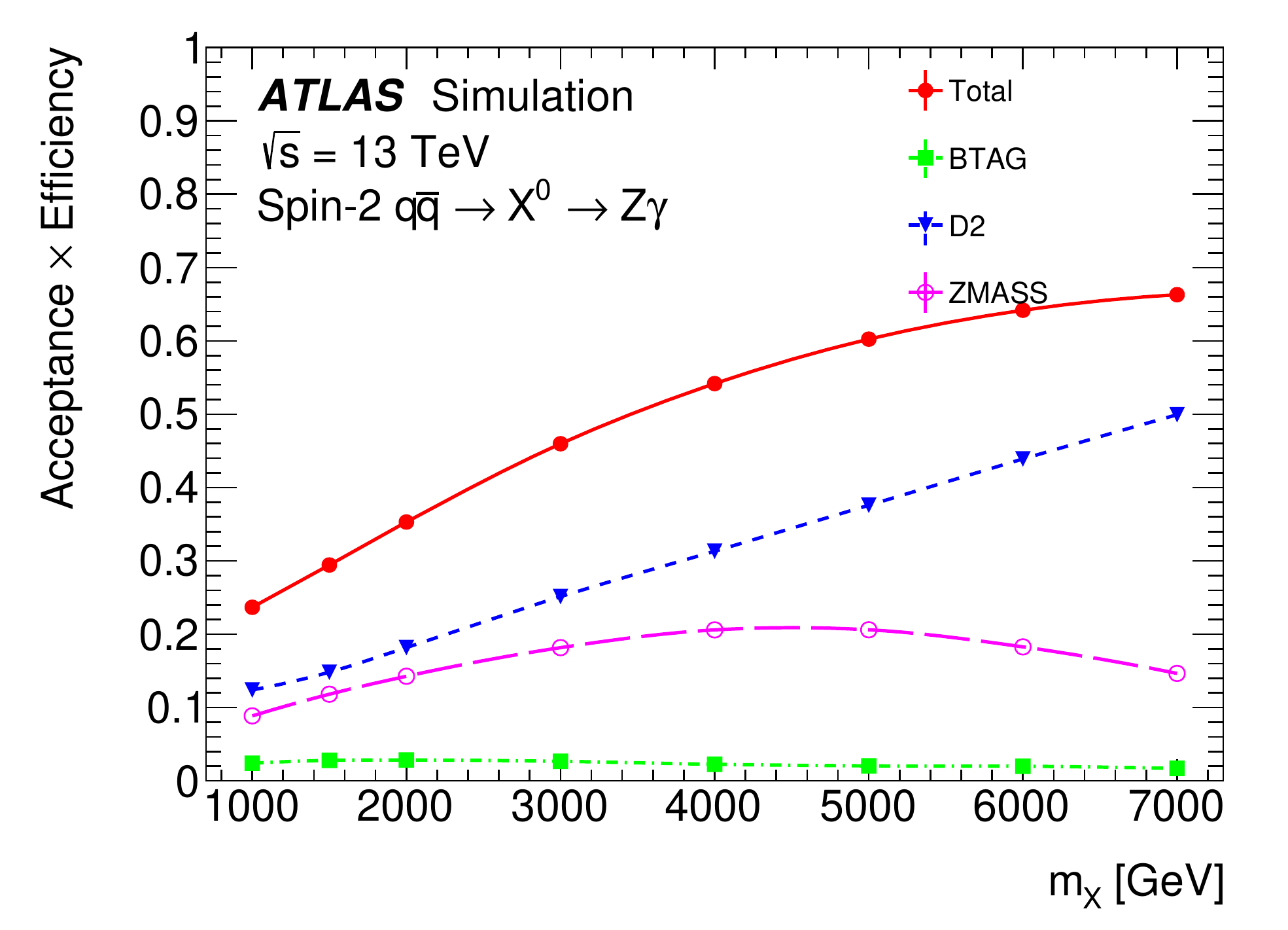} }
\subfigure[]{\includegraphics[width=0.44\textwidth]{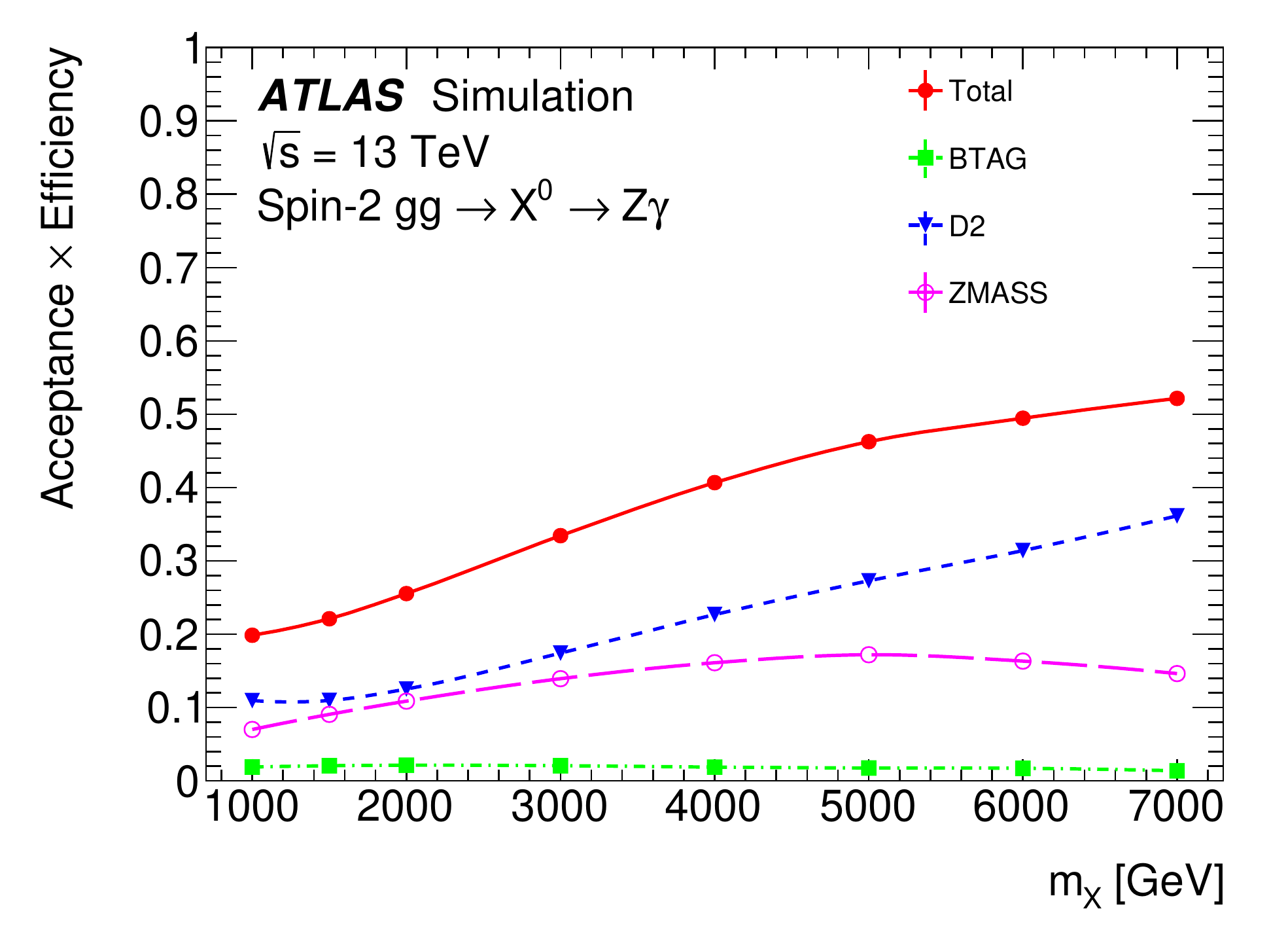} }
\caption{Total efficiencies for the selection of signal events after categorization
and application of the tighter photon \ETg{} selection used to optimize the
signal significance:
(a) \ggXZgJ{0},
(b) \qqXWgJ{1},
(c) \qqXZgJ{2},
and
(d) \ggXZgJ{2}.
In addition to the total efficiency, contributions to the signal selection
from each of the separate event categories are shown.
The efficiencies calculated from MC samples with $W/Z$ hadronic decays are shown as the points on each curve.
The line presents interpolated results.
}
\label{AbsoluteEfficiency}
\end{center}
\end{figure}

The rejection of the dominant $\gamma$+jet background varies strongly among the categories, being highest in those using jet substructure and mass information.
A further optimization of the signal sensitivity is implemented by varying the photon \ETg{} threshold as a function of the invariant mass $m_{J\gamma}$ of the photon and large-$R$ jet,
where the figure of merit is the statistical-only significance of the simulated BSM signal over the expected SM backgrounds, where the SM backgrounds are estimated from the simulated background samples described in Section~\ref{sec:mc}.
This photon \ETg{} optimization is done separately for each of the event categories, taking advantage of the large difference in photon and jet kinematics between signal and background.
The photon \ETg{} threshold increases with \mjg{}, varying from about 300 to 1200~\GeV.
This results in a small loss of signal efficiency, but a very large suppression of the SM backgrounds.
Figure~\ref{AbsoluteEfficiency} shows the total signal selection efficiencies after optimization of the photon \ETg{} thresholds, and also the contributions to the signal selection from each of the individual categories.
The BTAG category has the lowest efficiency but the highest signal purity.
The spin-2 $Z\gamma$ channel with $gg$ production mode has a different $X$ boson polarization than the $q\bar{q}$ production mode, leading to a longer lower tail in the photon and jet \pT{} distributions, and wider pseudorapidity distributions, and therefore a lower baseline selection efficiency.
For signals with a resonance mass above 4~\TeV, the applied $D_{2}$ requirement is relatively loose, which results in most signal events entering the D2 category and the W/ZMASS selection appearing to lose efficiency.
The signal selection efficiencies increase with the mass $m_{X}$, ranging from about 20\% at the lowest mass to about 60\% at 6.8~\TeV.
 
\section{Signal and background modelling}
\label{sec:signal_bkg_model}

 
The search for BSM boson signals is carried out by inspecting the invariant mass distribution of the highest-$\ET$ photon and large-$R$ jet identified in each event.
The distribution of $m_{J\gamma}$ from SM backgrounds falls smoothly over the mass range 1.0 to 6.8~\TeV\ used in this search.
The presence of a boson $X^{\pm}$$\to$ $W^{\pm}\gamma$  or $X^0$$\to$ $Z\gamma$ would therefore appear in the data as an excess of events above the
background $m_{J\gamma}$ distribution in a relatively narrow mass region around $m_{X}$.
The search sensitivity is quantified
by fitting the data with the sum of the SM background plus a signal that is parameterized from simulations of the various production modes described in Section~\ref{sec:mc}.
The functional forms chosen for the background and signal are described below, and the fitting  procedure used to search for signals is presented in Section~\ref{sec:statisticalTreatment}.
 
\subsection{SM background modelling}
 
The SM background is dominated by $\gamma$+jet events. In the D2 (W/ZMASS) event category,
the production of a photon in association with light-flavour jets and $c$-jets contributes about 92$\%$ (96$\%$) of the SM background,
while in the BTAG category the contribution from SM $\gamma$+$b$-jet events is about 88$\%$. The next highest background contribution comes from SM $W\gamma$ and $Z\gamma$ production with the $W/Z$ bosons decaying
hadronically. The contribution from SM $t\bar{t}$+$\gamma$ production is found to be negligible compared to other SM backgrounds after
the final event selection. Contributions from events with photons misidentified as jets are found to be small and not significant in changing the background shape from the dominant $\gamma$+jet backgrounds.
 
The $m_{J\gamma}$ distribution of the background is parameterized with a function that is flexible enough to accommodate the
background shape in each of the four event categories used in the signal search.
The function chosen to model the background is taken from Ref.~\cite{dijetfunction}, and is described
by Eq.~(\ref{back:eqn:background}):
\begin{equation}
\mathcal{B}(m_{J\gamma};\bm{p}) = (1-x)^{p_1}x^{p_2+p_3 \log(x)},
\label{back:eqn:background}
\end{equation}
where $x = m_{J\gamma}/\sqrt{s}$, and $\bm{p}$= ($p_1, p_2, p_3$) is a vector of parameters used to control the shape of the distribution.
The ability of the function to describe backgrounds is tested using $m_{J\gamma}$ distributions from MC simulations which have about five times the number of data events in the signal region. The number of parameters $p_i$ is determined by testing the
ability of each function to fit these background $m_{J\gamma}$ distributions over the mass range used for each category.
The determination of the number of parameters also includes studies of fits of the background-only mass distributions to a signal-plus-background hypothesis in order to quantify any \enquote{spurious signal} ($N^\text{SS}$) resulting from the parameterization with the procedure documented in Ref.~\cite{HIGG-2013-08}.
The number of fit parameters that minimizes the spurious signal is chosen. With this criterion, the number of fit parameters is two or three depending on the category and signal model.
The spurious signal is then included as a systematic uncertainty in the fitted signal yield associated with the background fit function, and included in the statistical treatment used for the signal search.
The choice of functional form and the spurious signal obtained from MC simulated samples are validated with data in a control region.
The control region (CR) events are selected with the photon required to be in the forward pseudorapidity region 1.52 $<$ $|\eta^\gamma|$$<$ 2.37.
This CR is found to have a small signal leakage which varies from 2\% to 17\% depending on the signal type and the resonance mass.
This validation process confirms that the chosen functional form is flexible enough to model the $m_{J\gamma}$ distribution in data.
 
\subsection{BSM signal modelling}
\label{sec:signalfitting}
 
The distribution of $m_{J\gamma}$ for a given BSM boson mass is generated with a natural width that is much smaller than the experimental resolution.
These MC events are passed through a full detector simulation and selected in the same way as data events.
The signal $m_{J\gamma}$ distribution is modelled with a double-sided Crystal Ball (DSCB) function~\cite{HIGG-2016-08}.
This function is found to be the best model to describe the peak and the long tails of the signal distribution.
It is described by Eq.~(\ref{eqn:sig_model_zg}):
 
\begin{equation}
\label{eqn:sig_model_zg}
\begin{aligned}
\mathcal{S}(\mjg; & N,\mu,\sigma,\alpha_1,n_1,\alpha_2,n_2) \\
= N \cdot &
\begin{cases}
\left( \frac{n_1}{|\alpha_1|} \right)^{n_1} \exp\left(- \frac{|\alpha_1|^2}{2}\right) \left( \frac{n_1}{|\alpha_1|} -|\alpha_1| - \frac{
\mjg - \mu}{\sigma} \right)^{-n_1}      & \quad \frac{\mjg - \mu}{\sigma} \leq - \alpha_1\\
\exp\left(- \frac{(\mjg - \mu)^2}{2\sigma^2}\right) & \quad -\alpha_1 < \frac{\mjg - \mu}{\sigma} \leq \alpha_2 \\
\left( \frac{n_2}{|\alpha_2|} \right)^{n_2} \exp\left(- \frac{|\alpha_2|^2}{2}\right) \left( \frac{n_2}{|\alpha_2|} -|\alpha_2| + \frac{\mjg - \mu}{\sigma} \right)^{-n_2}      & \quad \alpha_2 < \frac{\mjg - \mu}{\sigma}. \\
\end{cases}
\end{aligned}
\end{equation}
 
The DSCB function includes a central Gaussian core, to model the experimental resolution of the signal, with tails parameterized with power-law functions above and below the peak.
The Gaussian core has a mean $\mu$ and width $\sigma$, while the low (high) $m_{J\gamma}$ tail is fitted using the parameters $\alpha_1$ ($\alpha_2$) and $n_1$ ($n_2$), with all the parameters
constrained to be positive in the fit.
 
This signal model is used to fit the $m_{J\gamma}$ distribution generated from the four signal hypotheses at masses
ranging from 1.0 to 7.0~\TeV\ in steps of 1.0~\TeV, with one additional mass point at 1.5~\TeV.
A linear interpolation between adjacent mass points is performed for each of the fit parameters to obtain the signal shapes at intermediate mass values.
The width of the central core grows linearly from a $\sigma$ of about 30 to 120~\GeV\ as the boson mass increases from 1.0 to 7.0~\TeV.


\section{Systematic uncertainties}

The systematic uncertainties considered in this analysis come from the background estimation, the signal prediction and the detector performance.
The effects of these systematic uncertainties are parameterized according to their impact on the signal efficiency, the signal shape peak position and the core width of the signal shape.
All these uncertainties are included in the statistical procedure when fitting the signal-plus-background model to the data.
 
The potential bias from the background fit function describing the data $m_{J\gamma}$ distribution is evaluated using the spurious-signal test described in Section~\ref{sec:signal_bkg_model}.
A spurious signal is treated as a systematic uncertainty arising from the choice of background parameterization and only affects the signal yield during the fitting procedure.
Assuming there is no signal in the data, the impact of spurious-signal uncertainties when setting cross-section limits decreases from 20\% to a negligible value with increasing resonance mass.
 
The uncertainty in the luminosity determination affects the signal yield prediction.
The integrated luminosity is measured using the LUCID-2 Cherenkov detector~\cite{Avoni_2018} and calibrated with a van der Meer scan following the methodology documented in Ref.~\cite{DAPR-2013-01}.
This results in a 1.7\% uncertainty in the 139~fb$^{-1}$ integrated luminosity collected during the 2015--2018 data-taking period.
 
The uncertainty in the modelling of inelastic $pp$ pile-up collisions overlaid on the simulation introduces a 2\% uncertainty in the signal detection efficiency.

The uncertainty in the photon energy measurement affects the signal selection efficiency and the shape of the invariant mass $m_{J\gamma}$ distribution.
The photon energy is calibrated using the method described in Ref.~\cite{EGAM-2018-01}.
Various sources of uncertainty contribute to the measurement of the photon energy scale and the photon energy resolution.
The photon identification, isolation and trigger efficiencies are all measured from data following the method in Refs.~\cite{EGAM-2018-01,TRIG-2018-05}.
 
The uncertainty in large-$R$ jet energy and mass calibration also affects the signal selection efficiency and the $m_{J\gamma}$ shape.
The large-$R$ jet energy and mass are calibrated with the method described in Ref.~\cite{JETM-2018-02}.
The impact of the jet energy resolution uncertainty is estimated by applying Gaussian smearing to each jet so as to degrade the jet $\pt$ resolution
by 2\%~\cite{JETM-2018-02}.
To estimate the impact of the jet mass resolution (JMR) uncertainty, a similar method is used to degrade the JMR by 20\%.
Similarly, the effect of uncertainty in the $D_{2}$ resolution is estimated by degrading the $D_{2}$ resolution by 15\% with Gaussian smearing.
 
The uncertainty in the jet-flavour tagging efficiency measurement impacts both the signal selection efficiency and the $m_{J\gamma}$ distribution.
The jet-flavour tagging efficiency is measured in a data region enriched in $t\bar{t}$ events and compared with simulations to derive corrections~\cite{FTAG-2018-01}.
The uncertainties for high-\pT{} VR track-jets are extrapolated with simulated samples because there are too few events in data~\cite{ATL-PHYS-PUB-2021-003}.
The associated uncertainties are grouped into $b$-jet, $c$-jet and light-flavour jet components that are described by uncorrelated eigenvector variations.
 
The uncertainty in the signal selection efficiency due to the PDF set is evaluated using the eigenvector variations following the method in Ref.~\cite{LHAPDF}.
The uncertainty in the signal selection efficiency from the QCD scales is estimated from alternative samples with the renormalization scale (\muR) and factorization scale (\muF) varied by factors of 0.5 and 2 with the cases that differ by a factor of four being ignored.
The uncertainty in the signal selection efficiency from the parton shower is estimated from alternative \PYTHIA samples with different values of the A14 tune parameters, affecting the underlying events, initial/final-state radiation, multiple parton interactions and colour reconnection~\cite{ATL-PHYS-PUB-2014-021}.
 
The limited size of the generated signal samples introduces a systematic uncertainty in the signal parameterization with  analytic functions as described in Section~\ref{sec:signalfitting}.
Only the effect on the signal resolution is found to have a significant impact on the final result and is included in the statistical analysis as a systematic uncertainty.
 
Table~\ref{fig:systematic_impact} summarizes the main sources of signal uncertainty and their impact on the signal measurement.
The dominant uncertainties for the signal in this analysis come from jet mass scale, jet mass resolution and jet energy resolution.
 
\begin{table}
\begin{center}
\caption{
The impact of systematic uncertainties on the signal yield, signal peak position and signal peak resolution.
Presented numbers are derived before performing the statistical analysis.
A range of values shows the variation of the uncertainty across the $m_{X}$ range.
}
\label{fig:systematic_impact}
\begin{tabular}{c|c}
\hline\hline
Source of uncertainty & Impact on signal yield [\%] \\
\hline
Luminosity  &  1.7\; \\
Jet energy scale  & 1--7   \\
Jet mass scale &  \phantom{0}1--20 \\
Jet mass resolution  & \phantom{0}2--12  \\
Jet $D_{2}$ resolution &  2\phantom{.0\;} \\
Photon energy scale  &  0.2\:  \\
Photon energy resolution  &  0.1\;  \\
Flavour tagging &  1--8  \\
Pile-up &  0--3 \\
PDF & \phantom{0}2--12 \\
QCD & 2 \\
Parton shower & 1--2 \\
\hline
& Impact on signal peak position [\%] \\
\hline
Jet energy scale  &  0--4  \\
Jet mass scale &  0--1 \\
Photon energy scale  &   0.4\; \\
\hline
& Impact on signal resolution [\%] \\
\hline
Jet energy scale  & 1--7  \\
Jet mass scale &  \phantom{0}0--11 \\
Jet energy resolution  & \phantom{0}5--20  \\
Photon energy scale  & 0.2--2\phantom{.0}  \\
Photon energy resolution  &  0.2--1.2  \\
Flavour tagging &  0.2--4\phantom{.0}  \\
Signal sample statistics & 1--6 \\
\hline\hline
\end{tabular}
\end{center}
\end{table}


\section{Statistical analysis}
\label{sec:statisticalTreatment}


The search for BSM resonance signals above a smoothly falling background $m_{J\gamma}$ mass distribution is carried
out with a statistical procedure based on an unbinned likelihood fit over the $m_{J\gamma}$ spectrum, implemented in a \texttt{RooFit}~\cite{roofit} and \texttt{RooStats}~\cite{roostats}  framework.
The likelihood function is defined as the product of several factors using a Poisson model for the observed event yield in each category.
This product includes probabilities for events distributed in $m_{J\gamma}$ as described by a model based on the sum of signal ($\mathcal{S}$) and background ($\mathcal{B}$) probability density functions described in Section~\ref{sec:signal_bkg_model} and probabilities for auxiliary measurements with their prior distributions ($\mathcal{G}$). This can be written as:
 
\begin{equation}\label{eq:LLHdef}
\begin{aligned}
\mathcal{L}(\bm{m}^{\mathrm{obs}}_{J\gamma}|\,\sigma_{\mathrm{had}},\bm{\theta},\bm{\theta}^{\mathrm{SS}},\bm{N}^{\mathrm{B}}, \bm{p})&=\prod_{c\in \mathbb{C}}\Biggl\{ \mathrm{Pois}(N_{c}^{\mathrm{obs}}|N_{c}^{\mathrm{S}}(\sigma_{\mathrm{had}},\bm{\theta})+N_{c}^{\mathrm{SS}}(\theta^{\mathrm{SS}}_{c}) +N_{c}^{\mathrm{B}}) \Biggr. \\
&\prod_{i=1}^{N_{c}^{\mathrm{obs}}}  \left[
\left(\frac{N_{c}^{\mathrm{S}}(\sigma_{\mathrm{had}},\bm{\theta})+N_{c}^{\mathrm{SS}}(\theta^{\mathrm{SS}}_{c})}{N_{c}^{\mathrm{S}}(\sigma_{\mathrm{had}},\bm{\theta}) +N_{c}^{\mathrm{SS}}(\theta^{\mathrm{SS}}_{c})+N_{c}^{\mathrm{B}}}\right)\mathcal{S}(\mjg^{c,i,\mathrm{obs}}|\,\bm{\theta})+ \right. \\
& \Biggl. \left. \left(\frac{N_{c}^{\mathrm{B}}}{N_{c}^{\mathrm{S}}(\sigma_{\mathrm{had}},\bm{\theta})+N_{c}^{\mathrm{SS}}(\theta^{\mathrm{SS}}_{c})+N_{c}^{\mathrm{B}}}\right)\mathcal{B}( \mjg^{c,i,\mathrm{obs}}|\,\bm{p}^{c})  \right] \Biggr\} \times \\
&\prod_{s\in \mathbb{S}}\mathcal{G}(0|\theta_{s},1) \prod_{c\in \mathbb{C}} \mathcal{G}(0|\theta^{\mathrm{SS}}_{c},1),
\end{aligned}
\end{equation}
where $\bm{m}^{\mathrm{obs}}_{J\gamma}=\{\mjg^{1,1,\mathrm{obs}},\cdots, \mjg^{c,i,\mathrm{obs}},\cdots\}$ is a set of observations of $\mjg$ in data, $c$ is the label of the various event categories and $i$ the index of events in each category.
The Poisson term for each category, $\mathrm{Pois}(N_{c}^{\mathrm{obs}}|N_{c}^{\mathrm{S}}(\sigma_{\mathrm{had}},\bm{\theta})+N_{c}^{\mathrm{SS}}+N_{c}^{\mathrm{B}})$, is defined according to observed data events in the signal region, $N_{c}^{\mathrm{obs}}$, and the expected signal-plus-background yield, which is a sum of the signal yield $N_{c}^{\mathrm{S}}(\sigma_{\mathrm{had}},\bm{\theta})$, the background yield $N_{c}^{\mathrm{B}}$, and the spurious signal $N_{c}^{\mathrm{SS}}$.
The signal yield $N_{c}^{\mathrm{S}}$ can be expanded as a function of the signal production cross-section $\sigma_{\mathrm{had}}$, which is the parameter of interest (POI) in the statistical analysis.
This cross-section $\sigma_{\mathrm{had}}$, as the abbreviation for $\sigma(pp\to X\to W/Z(\to \mathrm{hadrons})+\gamma)$, includes the production cross-section $\sigma(pp\to X)$ of the resonance and the branching fractions of $X\to W/Z+\gamma$ and $W/Z\to\mathrm{hadrons}$.
The experimental and theoretical uncertainties are described by the nuisance parameters (NPs) $\theta_{s}$ for each systematic uncertainty $s$ and shared among categories.
A collection of such nuisance parameters is written as $\bm{\theta}$. These nuisance parameters are constrained with a normal distribution $\mathcal{G}(0|\theta_{s},1)$.
The spurious-signal contribution $N_{c}^{\mathrm{SS}}$ is formalized as a function of the associated nuisance parameter $\theta^{\mathrm{SS}}_{c}$ for each category individually, with this NP following a normal distribution $\mathcal{G}(0|\theta^{\mathrm{SS}}_{c},1)$.
The collection of spurious-signal nuisance parameters is written as $\bm{\theta}^{\mathrm{SS}}$.
Both $\theta_{s}$ and $\theta^{\mathrm{SS}}_{c}$ can have an impact on the signal expectation ($N_{c}^{\mathrm{S}}+N_{c}^{\mathrm{SS}}$) of the fit model, while the parameter $\theta_{s}$ can also modify the signal shape.
The background shape parameters $\bm{p}^{c}=(p_{1}^{c}, p_{2}^{c}, p_{3}^{c})$ are allowed to float during the fit to data and are uncorrelated among categories.
The signal model $\mathcal{S}$ is fixed for each tested $m_{X}$ using the coefficients presented in Section~\ref{sec:signal_bkg_model}.
 
Both the signal and background yields are extracted by maximizing the likelihood as defined in Eq.~(\ref{eq:LLHdef}) for various hypothetical values of $m_{X}$.
The fit stability is checked with signal injection tests, and no significant bias is observed.
For each of these mass points, the $p$-value of the background-only hypothesis is calculated to test the compatibility of the background-only hypothesis and the data.
This is done with the profiled likelihood ratio (PLR) test statistic~\cite{Cowan:2010js}, which is defined as the ratio of the conditional maximum-likelihood value for a POI value of zero to the global maximum-likelihood value.
Its distribution in the low resonance mass region ($m_{X}< 4000\,\GeV$) is derived following the asymptotic approach as described in Ref.~\cite{Cowan:2010js}.
In the high resonance mass region ($m_{X}\ge 4000\,\GeV$), test statistic distributions are obtained with the pseudo-experiment sampling method.
The $p$-value reflects the possibility of background to produce a signal-like excess larger than that found in the fit to the data, which is reported as the significance according to the normal distribution.
Beside the significance, an exclusion of the signal model is derived and presented as the 95\% confidence level (CL) upper limit on the resonance production cross-section times branching fraction of $X\rightarrow W/Z+\gamma$ for hadronic decay of the $W/Z$ bosons.
Similar to the $p$-value, the upper limit is also calculated from PLR distributions but with a running POI value to indicate various signal cross-section hypotheses. The $\mathrm{CL_{s}}$ approach~\cite{CLs2,CLs1} is used for the limit calculation.
The limits are calculated in the low resonance mass regions at 20~\GeV\ steps and are based on the asymptotic approach.
In the high resonance mass region, limits are derived by using the pseudo-experiment sampling method.
To obtain smooth expected limit bands, the expected limits and the corresponding bands are calculated at 500~\GeV\ steps in the high resonance mass region while the observed ones are obtained at 100~\GeV\ steps.
Upper limits on $\sigma(pp\rightarrow X)\times B(X\to W/Z+\gamma)$ are derived by assuming the branching fractions of $W$ and $Z$ bosons to hadrons to be 67.41\%~\cite{pdg_physics_review} and 69.91\%~\cite{pdg_physics_review} respectively.


\section{Results}


Table~\ref{tab:data_yield} presents the observed number of events in different categories after the final event selection.
The yields quoted are for  $m_{J\gamma} \geq  800$~\GeV\ in the BTAG and D2 categories and for $m_{J\gamma} \geq 1000$~\GeV\ in the VMASS (ZMASS or WMASS) categories.
The BTAG categories are defined in the same way for the three $Z$ signal hypotheses, while for the D2 and VMASS categories the selection criteria for the photon and jet are chosen differently for each channel.
The latter optimizes the signal significance by exploiting differences in the $W/Z+\gamma$ production angular distributions and in the decays of the longitudinally polarized $W$ bosons and transversely polarized $Z$ bosons.

\begin{table}[tbph]
\begin{center}
\caption{Data yields in various categories defined for the four search channels.}
\label{tab:data_yield}
\begin{tabular}{l|c|c|c}
\hline
\hline
Channel & BTAG & D2 & VMASS \\
\hline
\ggXZgJC{0} & 436 & \phantom{1}5\,659	&  20\,728 \\
\ggXZgJC{2} & 436 & 10\,772  			&  32\,281 \\
\qqXZgJC{2} & 436 & \phantom{1}5\,618	&  18\,264 \\
\qqXWgJC{1}  & --- & \phantom{1}6\,373	&  25\,146 \\
\hline
\hline
\end{tabular}
\end{center}
\end{table}

\begin{figure}[h]
\begin{center}
\subfigure[]{\includegraphics[width=0.45\textwidth]{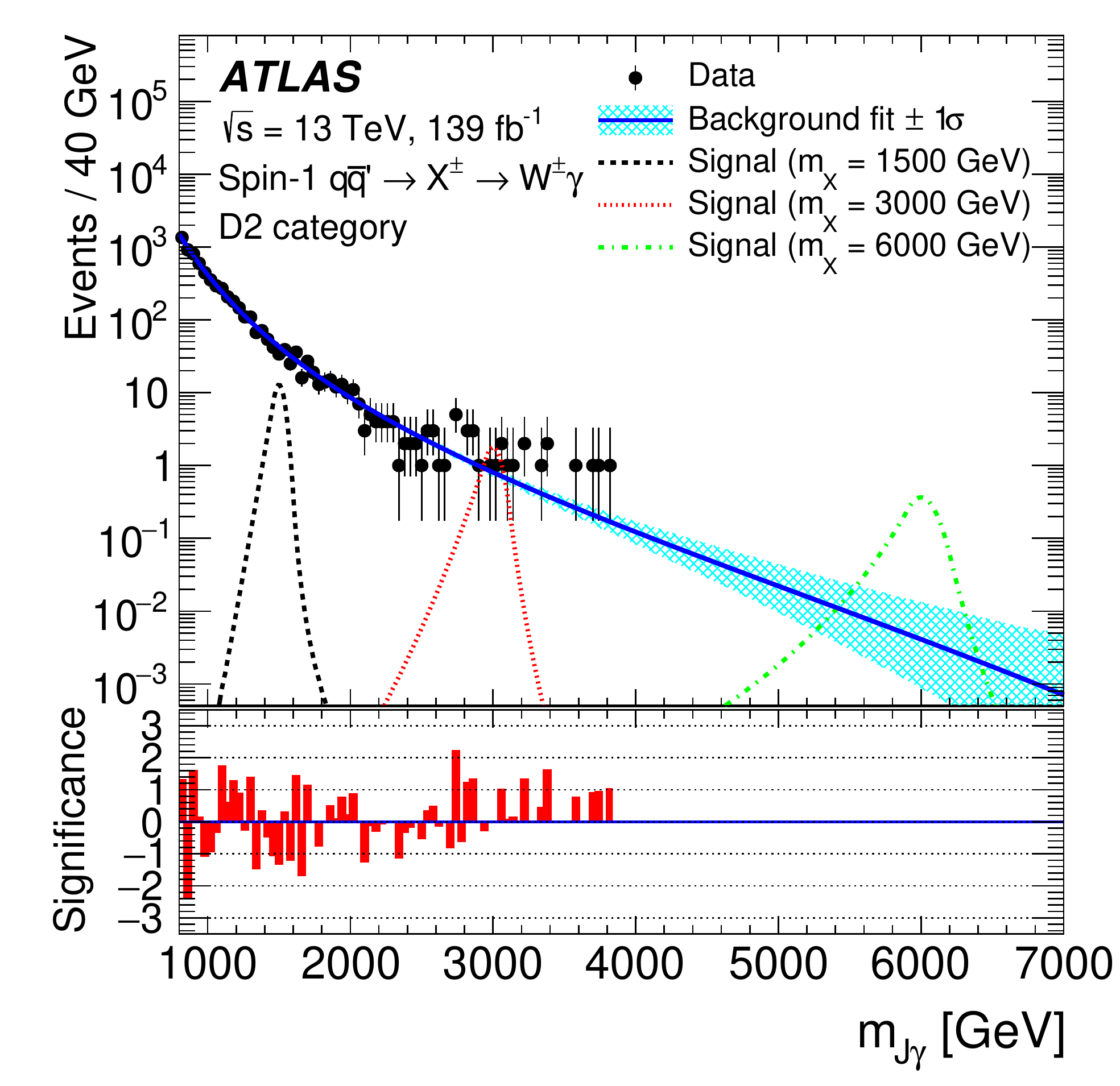}}
\subfigure[]{\includegraphics[width=0.45\textwidth]{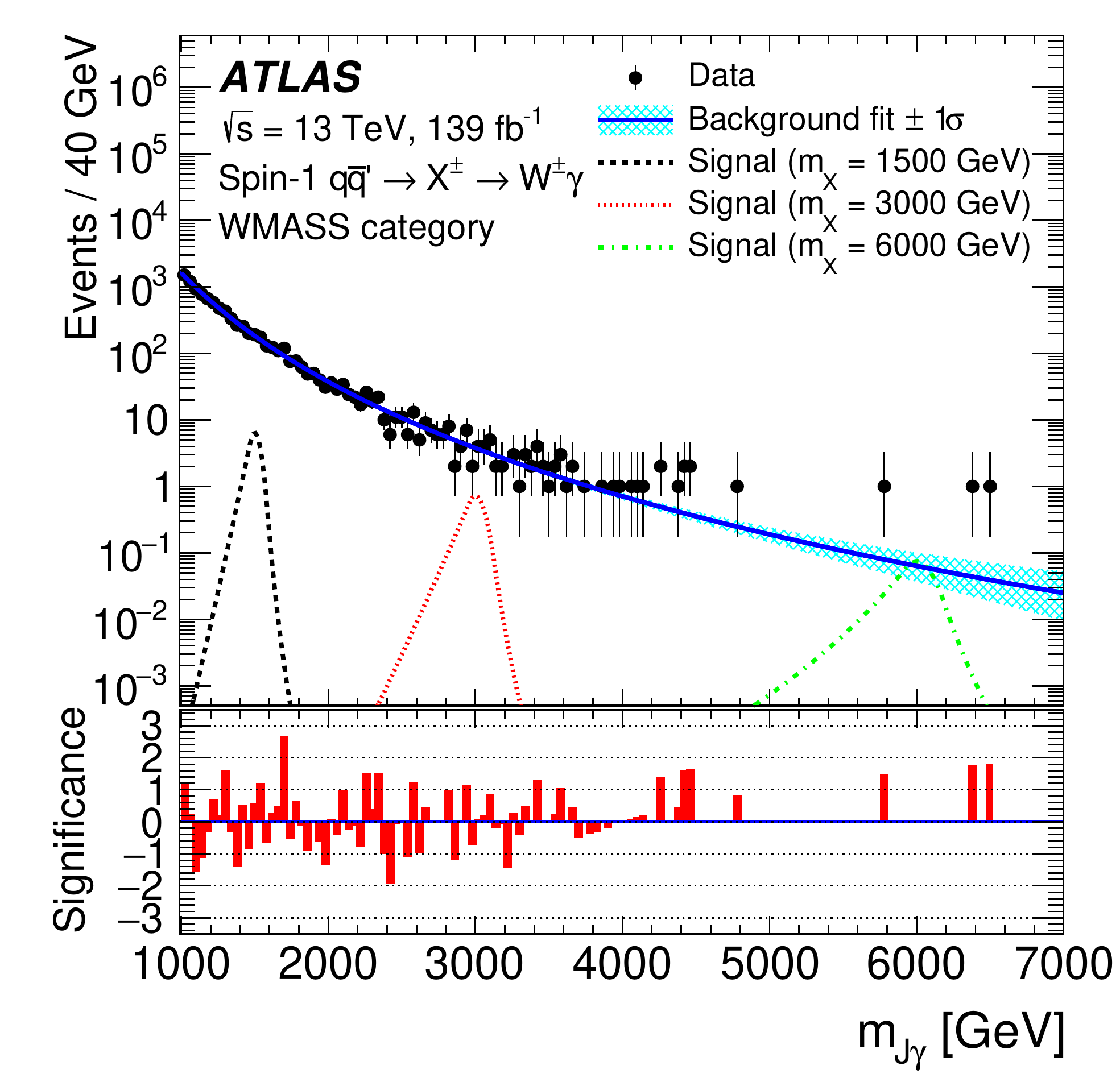}}
\end{center}
\caption{The $\mjg$ distributions of data events selected for the \qqXWgJ{1} search in the
(a) D2 and
(b) WMASS categories.
The background-only fit function shape is shown as the solid curve overlaid with a shaded band corresponding to statistical uncertainties in background parameters.
Various signal shapes with cross-sections corresponding to expected limits are shown as dashed lines.
The bottom panel presents the binned local significance (filled bars) from a comparison of the data with the background fit using a Poisson model~\cite{Choudalakis_2012}.
}
\label{fig:s1Wg_data_dist}
\end{figure}

\begin{figure}[h]
\begin{center}
\subfigure[]{\includegraphics[width=0.45\textwidth]{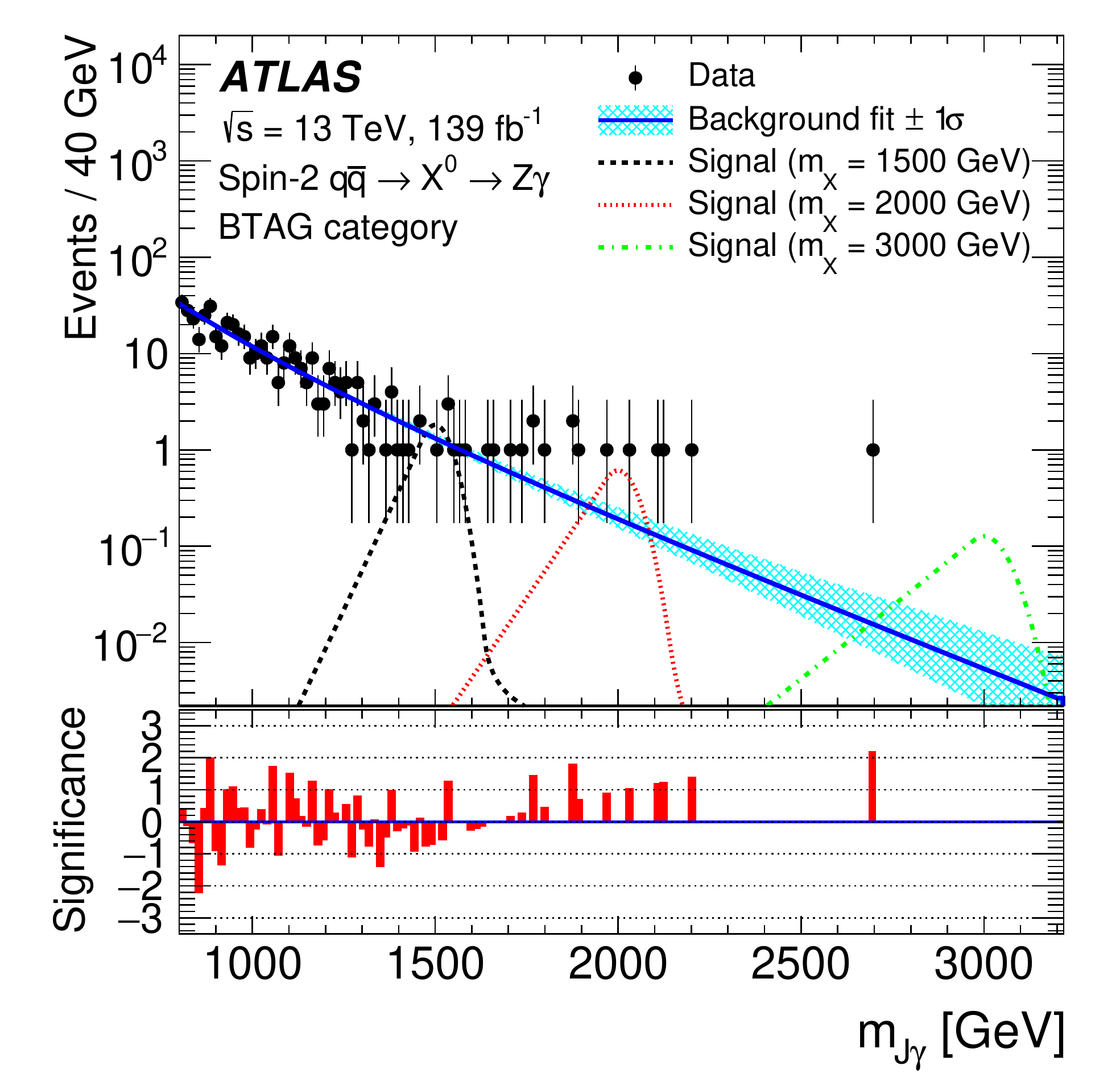}}
\subfigure[]{\includegraphics[width=0.45\textwidth]{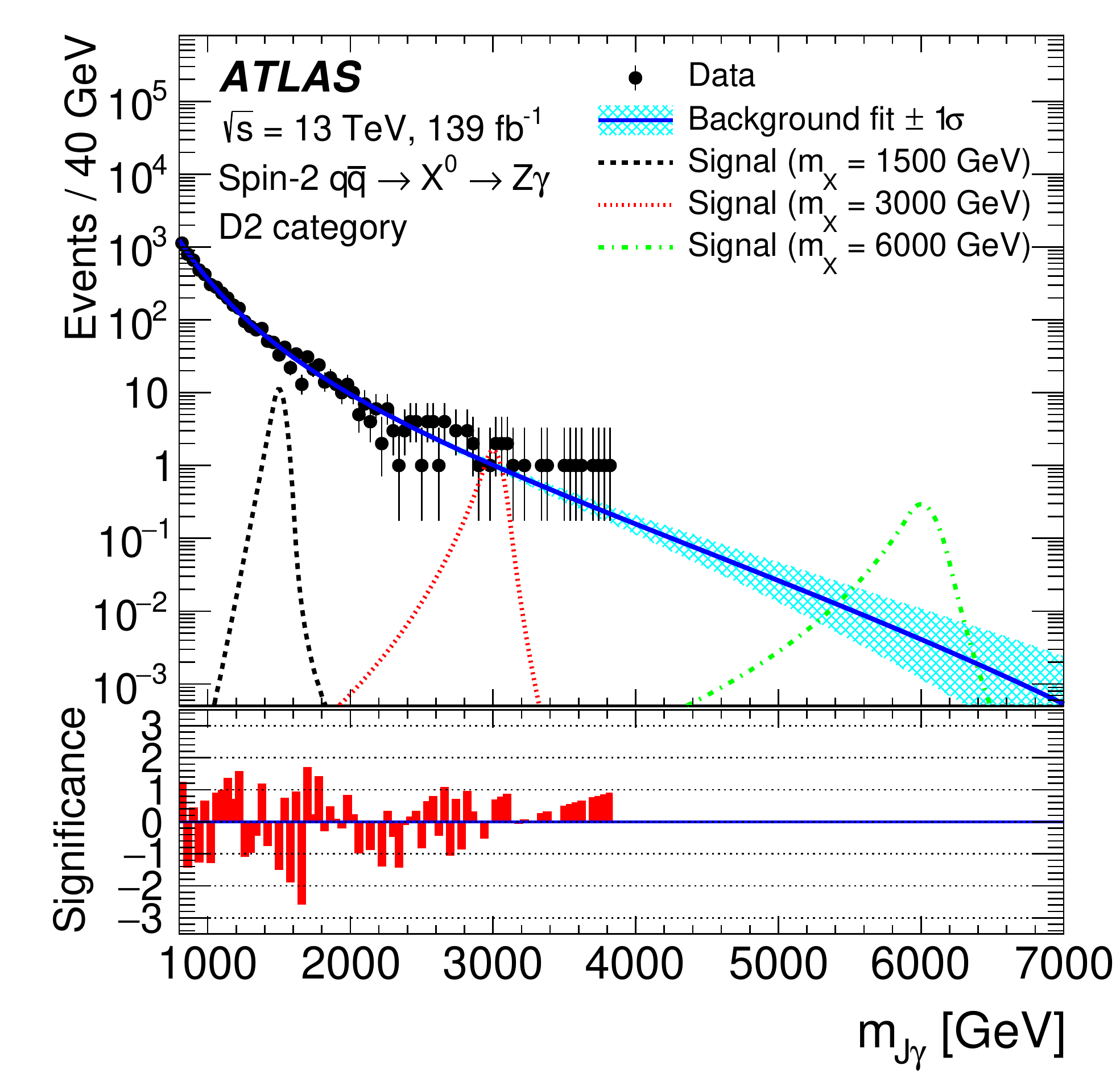}}
\subfigure[]{\includegraphics[width=0.45\textwidth]{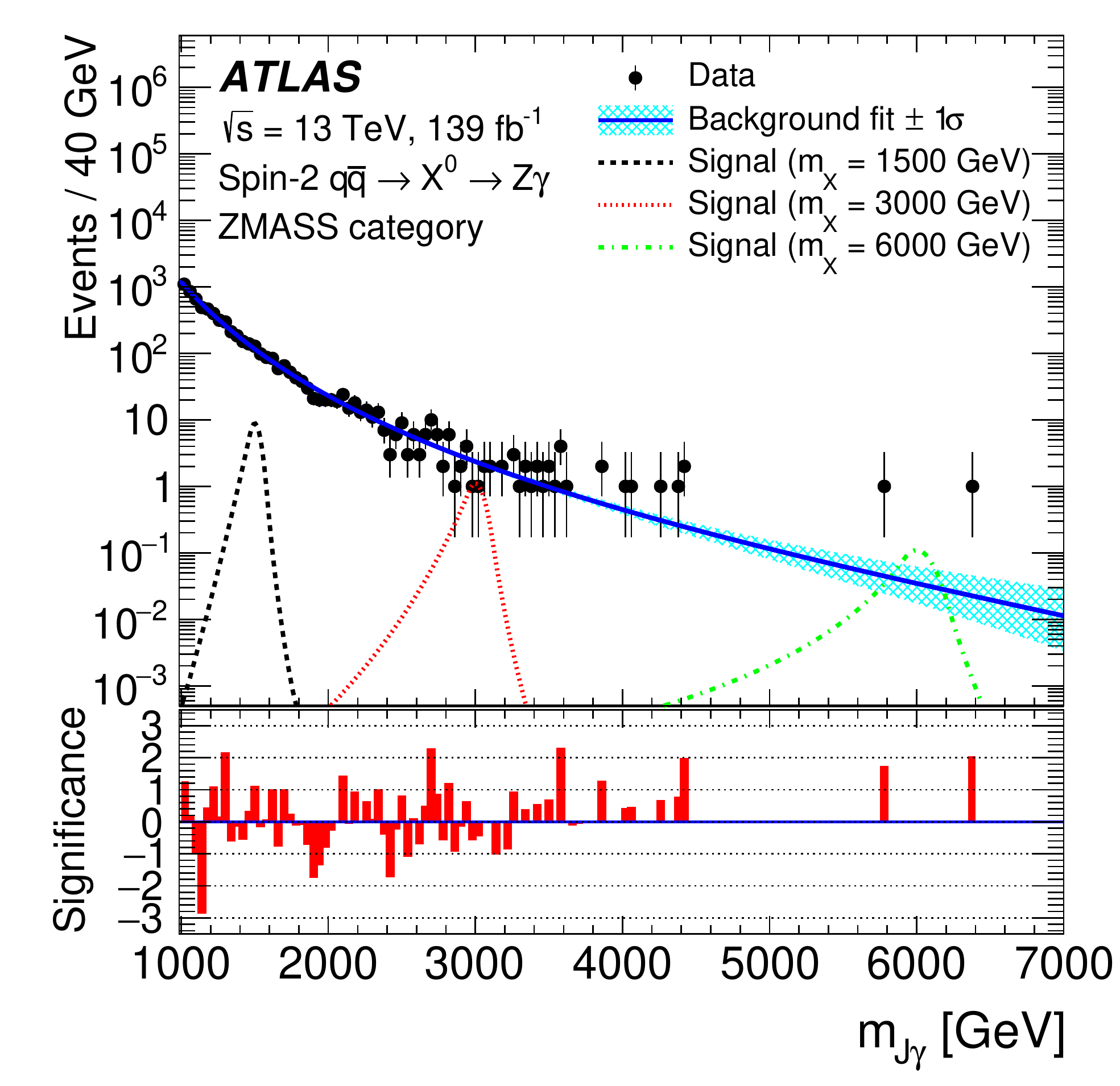}}
\end{center}
\caption{The $\mjg$ distributions of data events selected for the \qqXZgJ{2} search in the
(a) BTAG,
(b) D2, and
(c) ZMASS categories.
The background-only fit function shape is shown as the solid curve overlaid with a shaded band corresponding to statistical uncertainties in background parameters.
Various signal shapes with cross-sections corresponding to expected limits obtained in this analysis are shown as dashed lines.
The bottom panel presents the binned local significance (filled bars) from a comparison of the data with the background fit using a Poisson model~\cite{Choudalakis_2012}.
}
\label{fig:s2qqZg_data_dist}
\end{figure}

The $m_{J\gamma}$  distributions in different categories are shown in Figures~\ref{fig:s1Wg_data_dist}--\ref{fig:s2qqZg_data_dist} for the four signal channels.
The background-only fit result is shown as the solid curve overlaid with a shaded band corresponding to statistical uncertainties in background parameters.
Various signal mass hypotheses are also plotted, where the signal cross-sections correspond to the expected upper limits obtained in this analysis.
For the BTAG category, the fit range is limited to below 3200~\GeV\ due to the significant loss of sensitivity because of the decrease in $b$-tagging efficiency beyond that range, while for other categories the fit upper boundary is 7000~\GeV.
The bottom panel presents the binned local significance (filled bars) from a comparison of the data with the background fit using a Poisson model~\cite{Choudalakis_2012}.
The background-only model fits the data well, with most of the deviations of the data from the background-only hypothesis having a local significance below two standard deviations.
 
\begin{figure}[th]
\begin{center}
\subfigure[]{\includegraphics[width=0.45\textwidth]{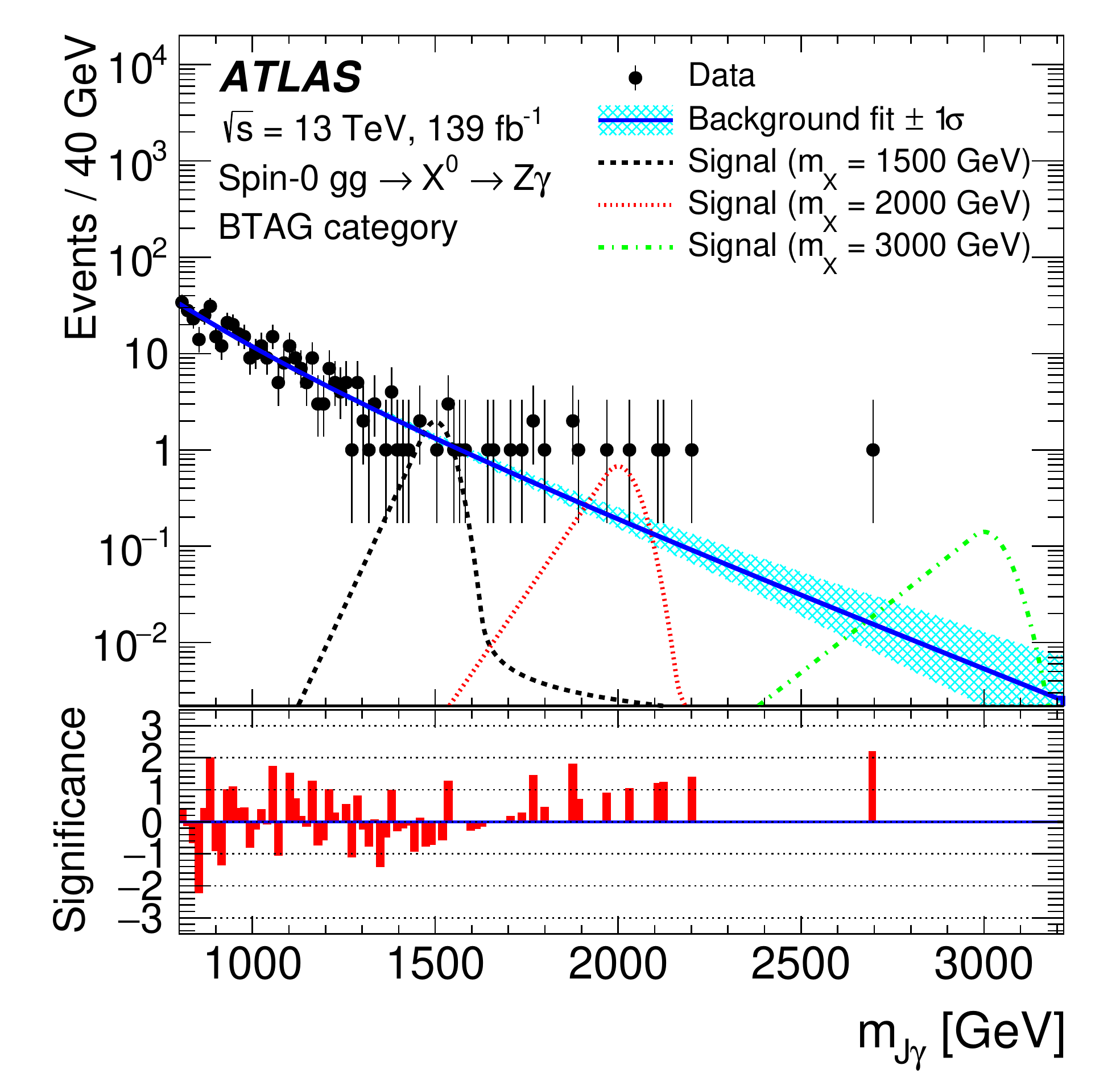}}
\subfigure[]{\includegraphics[width=0.45\textwidth]{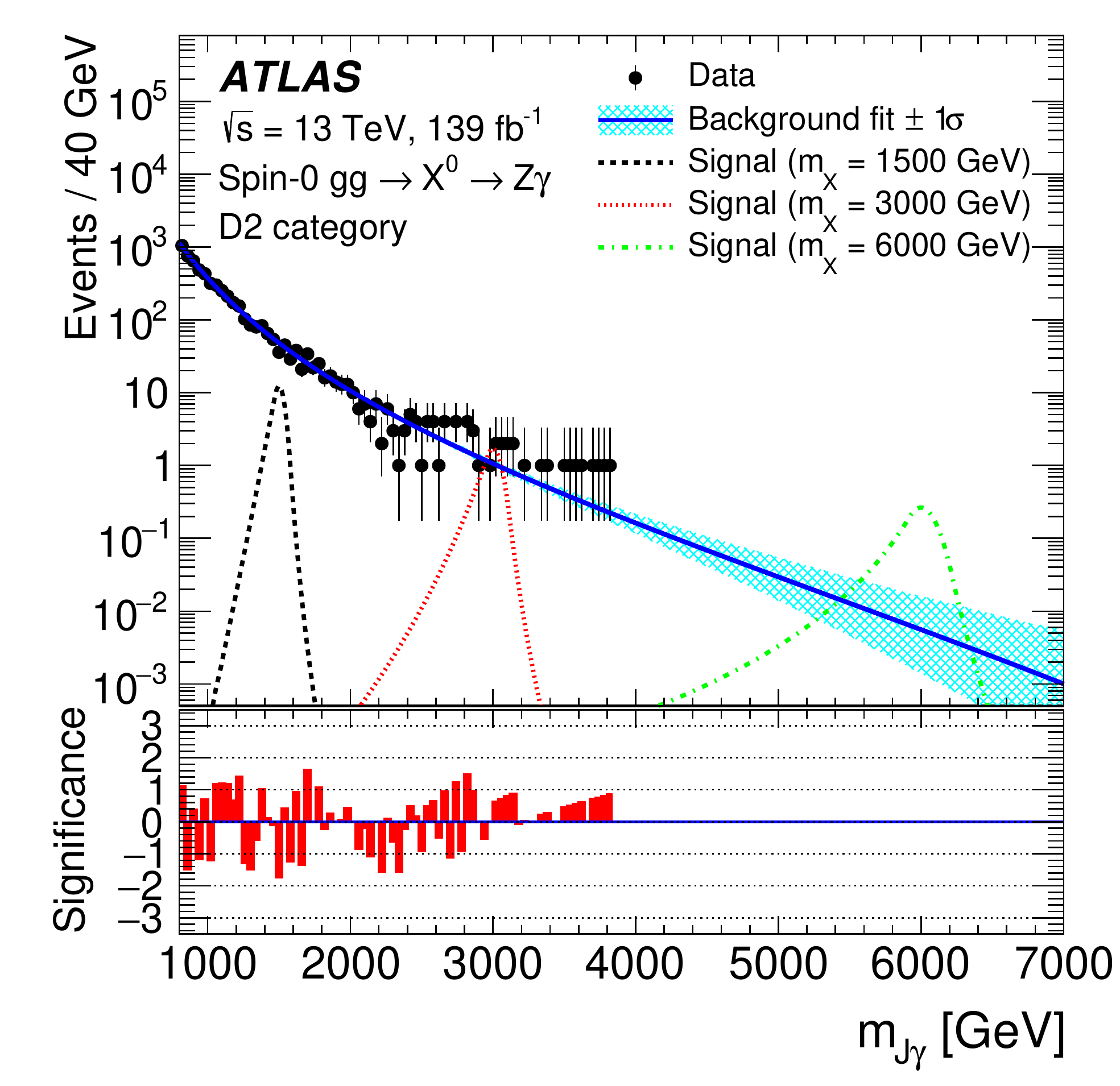}}
\subfigure[]{\includegraphics[width=0.45\textwidth]{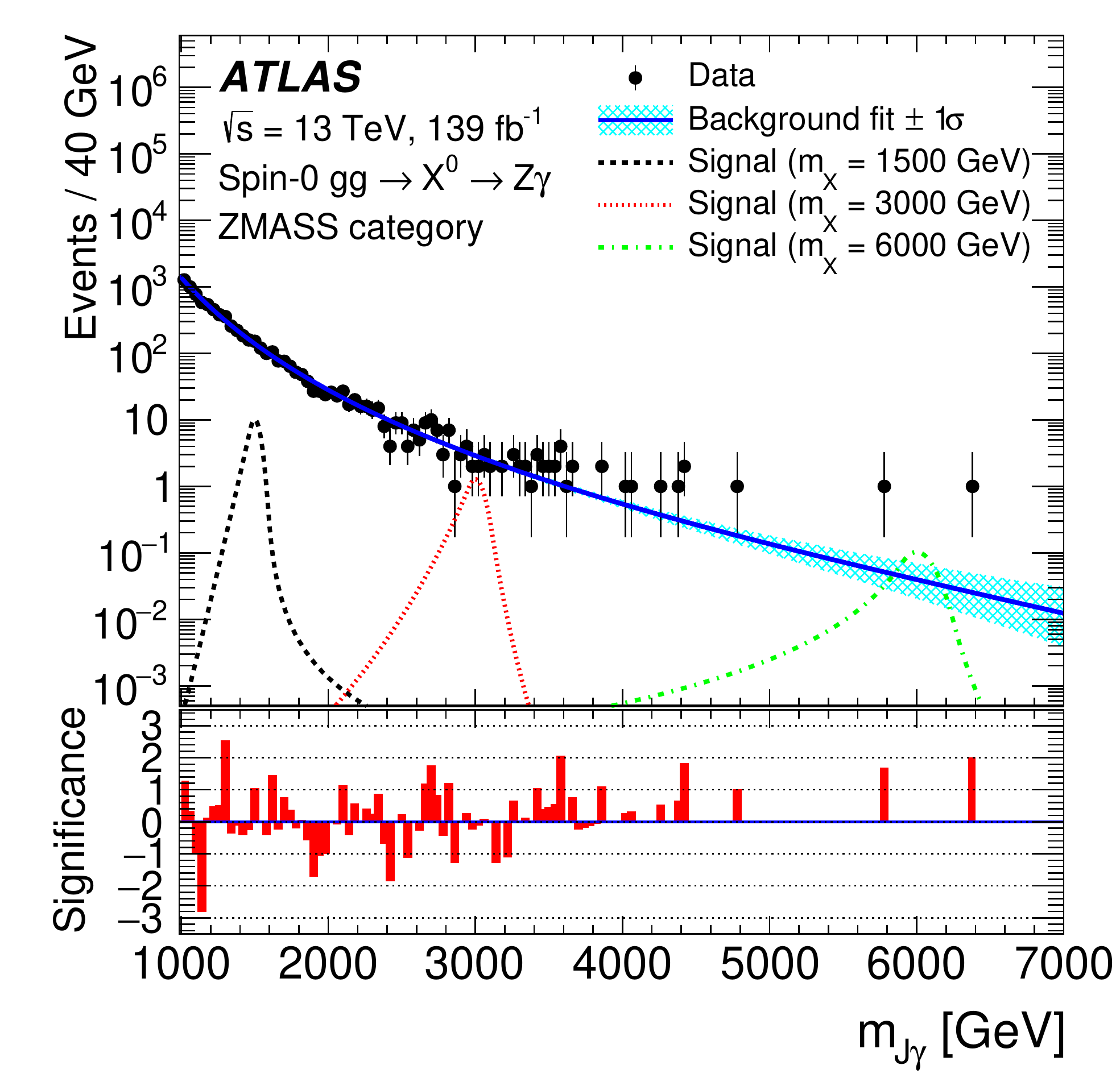}}
\end{center}
\caption{The $\mjg$ distributions of data events selected for the \ggXZgJ{0} search in the
(a) BTAG,
(b) D2, and
(c) ZMASS categories.
The background-only fit function shape is shown as the solid curve overlaid with a shaded band corresponding to statistical uncertainties in background parameters.
Various signal shapes with cross-sections corresponding to expected limits are shown as dashed lines.
The bottom panel presents the binned local significance (filled bars) from a comparison of the data with the background fit using a Poisson model~\cite{Choudalakis_2012}.
}
\end{figure}
 
\begin{figure}[h]
\begin{center}
\subfigure[]{\includegraphics[width=0.45\textwidth]{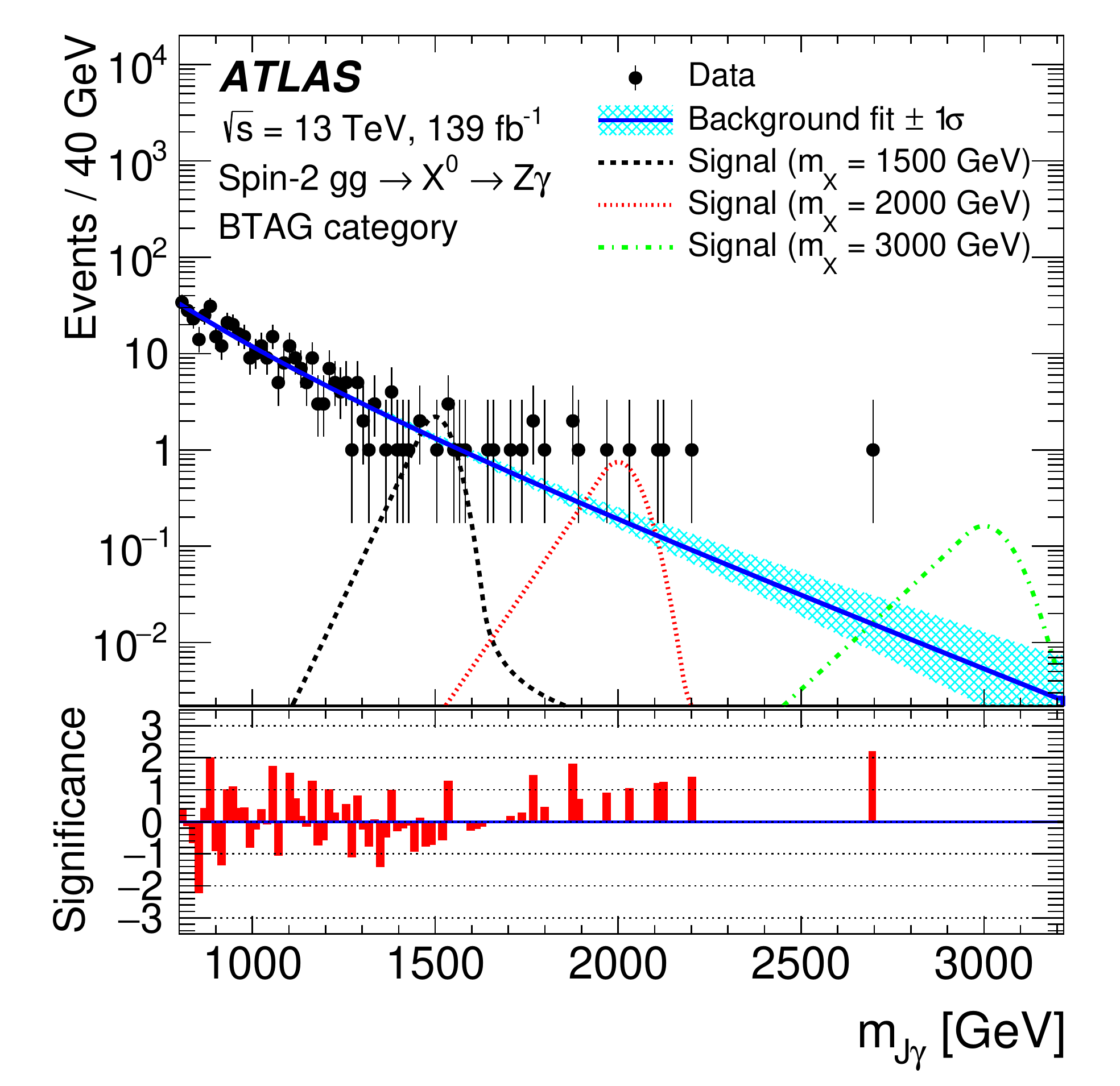}}
\subfigure[]{\includegraphics[width=0.45\textwidth]{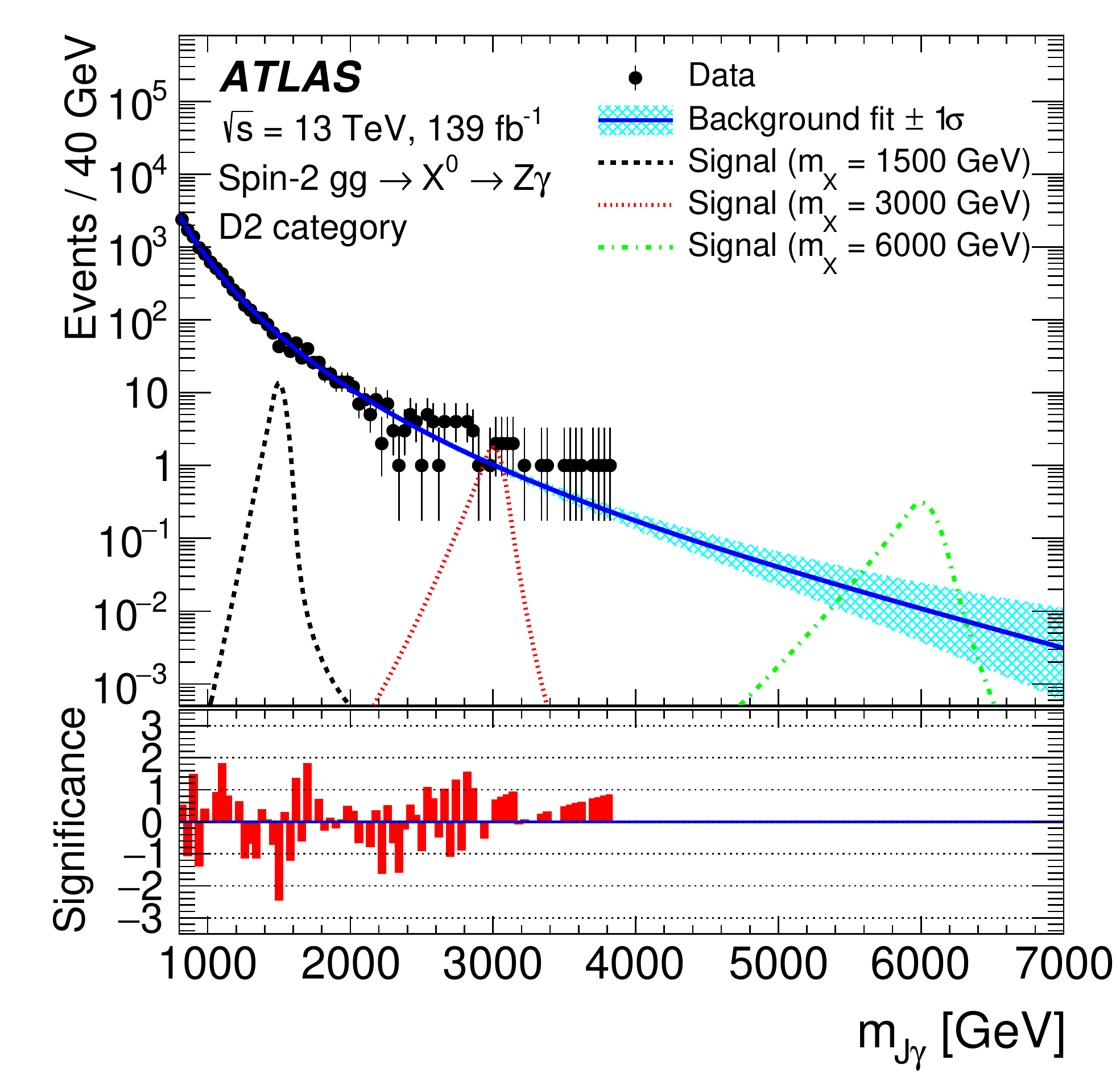}}
\subfigure[]{\includegraphics[width=0.45\textwidth]{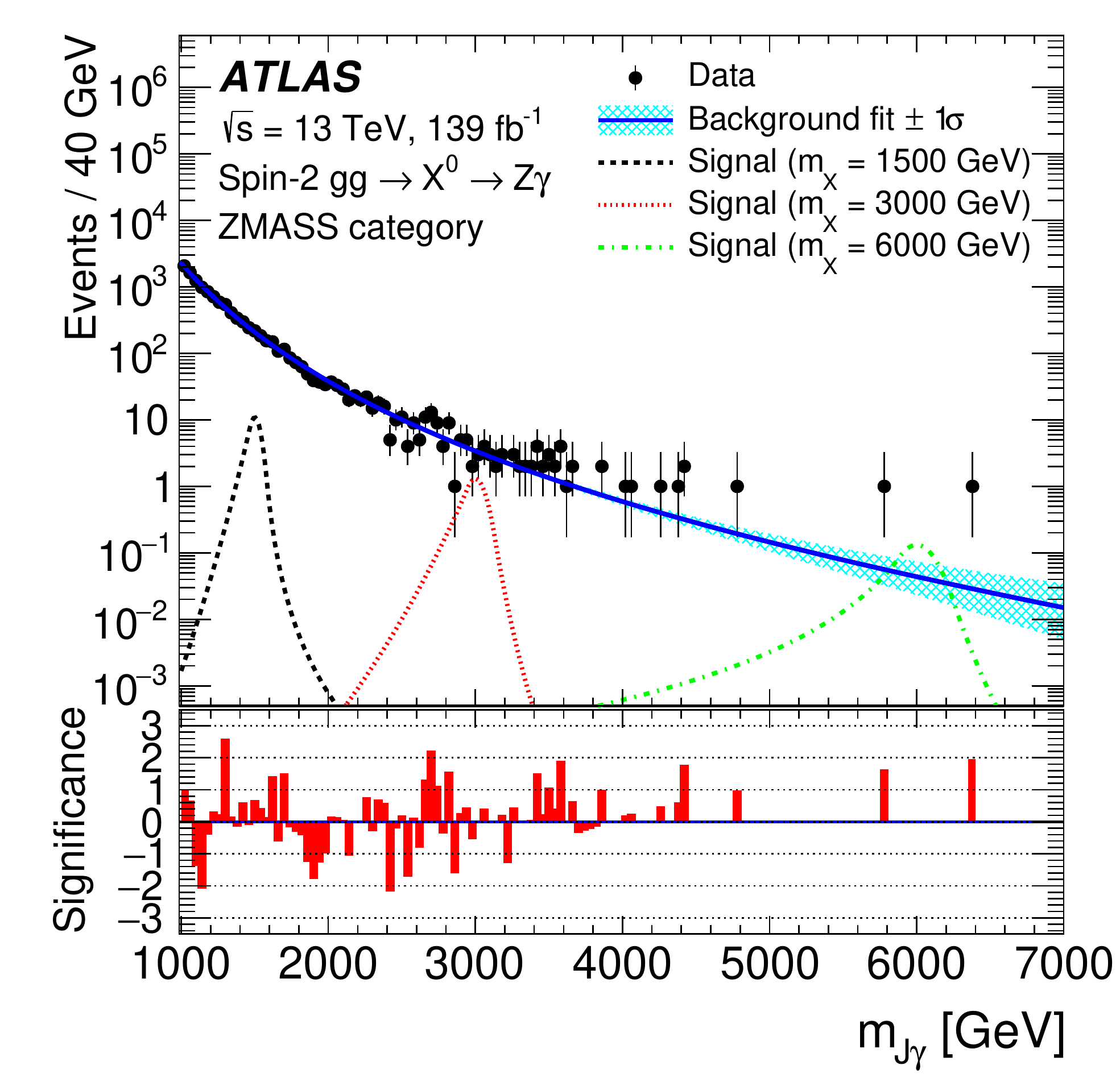}}
\end{center}
\caption{The $\mjg$ distributions of data events selected for the \ggXZgJ{2} search in the
(a) BTAG,
(b) D2, and
(c) ZMASS categories.
The background-only fit function shape is shown as the solid curve overlaid with a shaded band corresponding to statistical uncertainties in background parameters.
Various signal shapes with cross-sections corresponding to expected limits are shown as dashed lines.
The bottom panel presents the binned local significance (filled bars) from a comparison of the data with the background fit using a Poisson model~\cite{Choudalakis_2012}.
}
\end{figure}
 
\FloatBarrier
 
\begin{figure}[th]
\begin{center}
\subfigure[]{\includegraphics[width=0.45\textwidth]{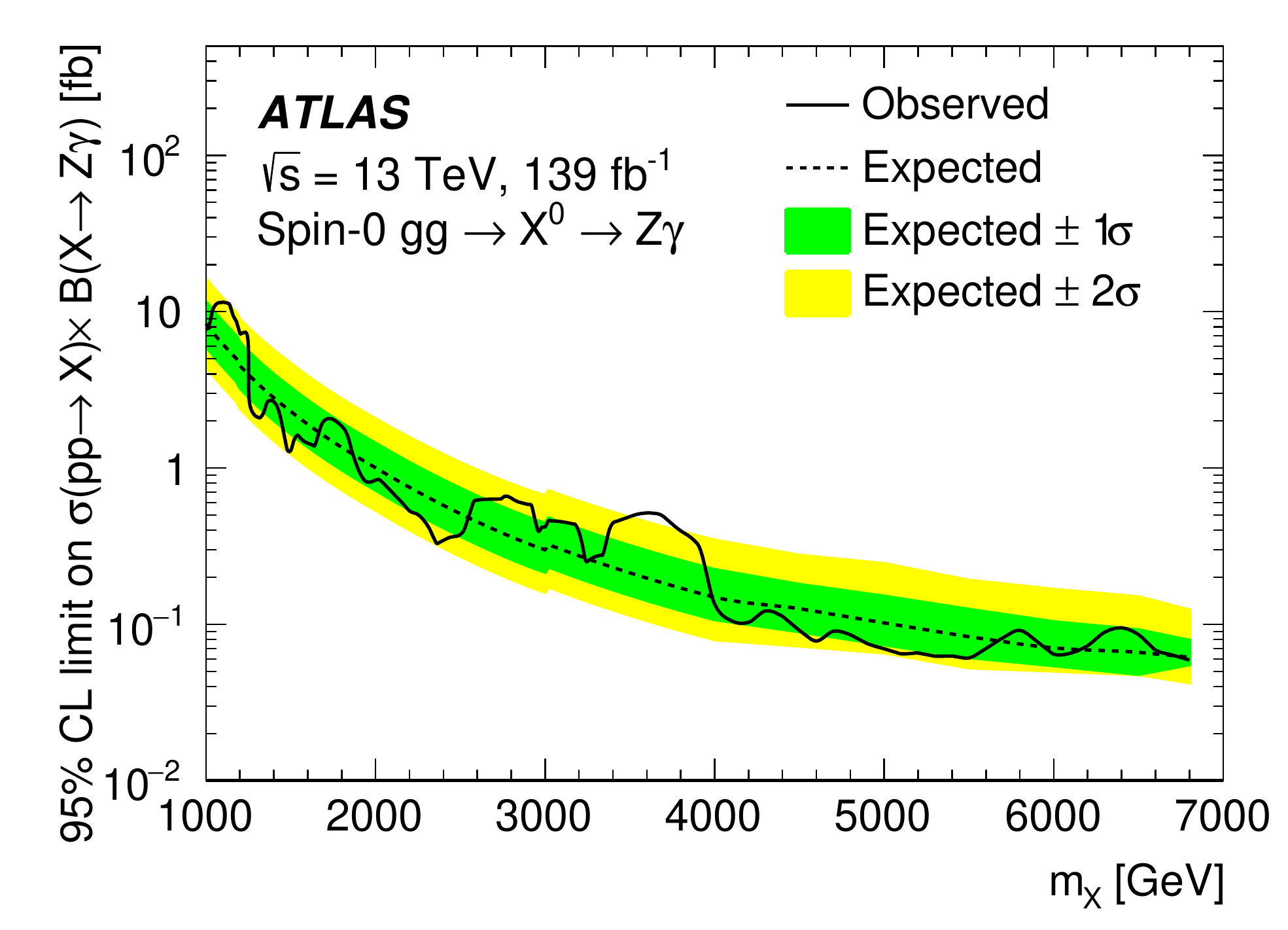}}
\subfigure[]{\includegraphics[width=0.45\textwidth]{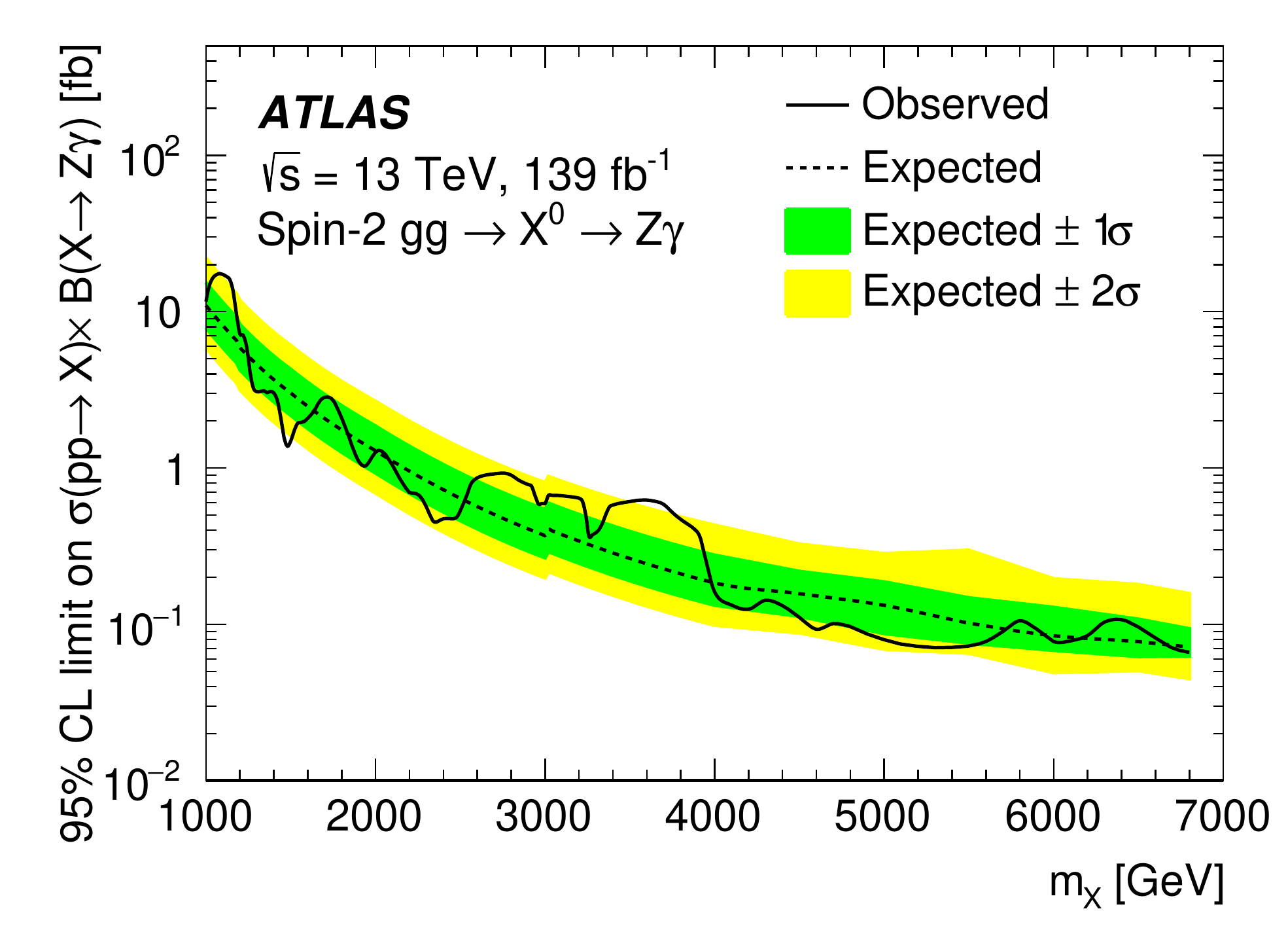}}
\subfigure[]{\includegraphics[width=0.45\textwidth]{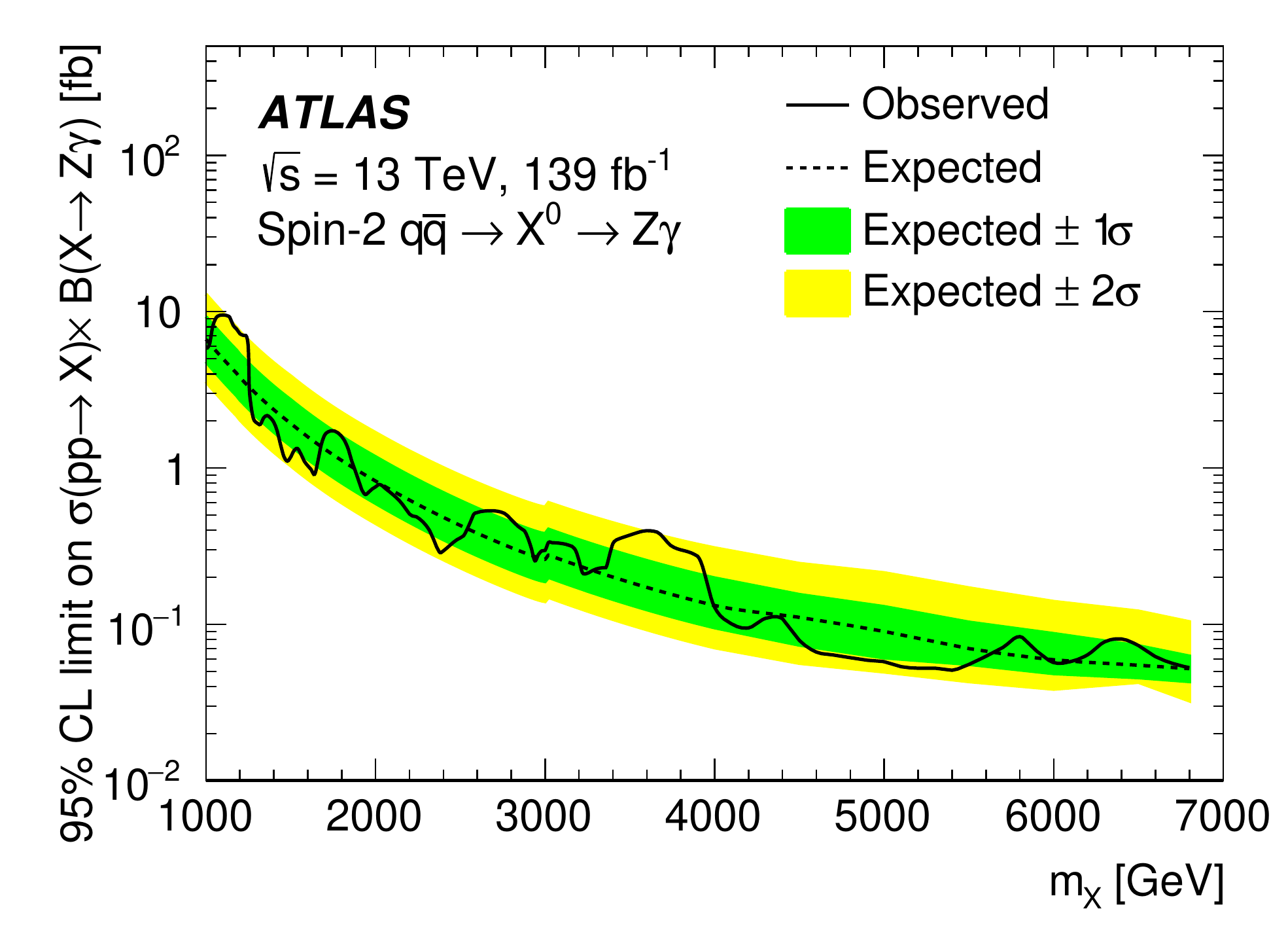}}
\subfigure[]{\includegraphics[width=0.45\textwidth]{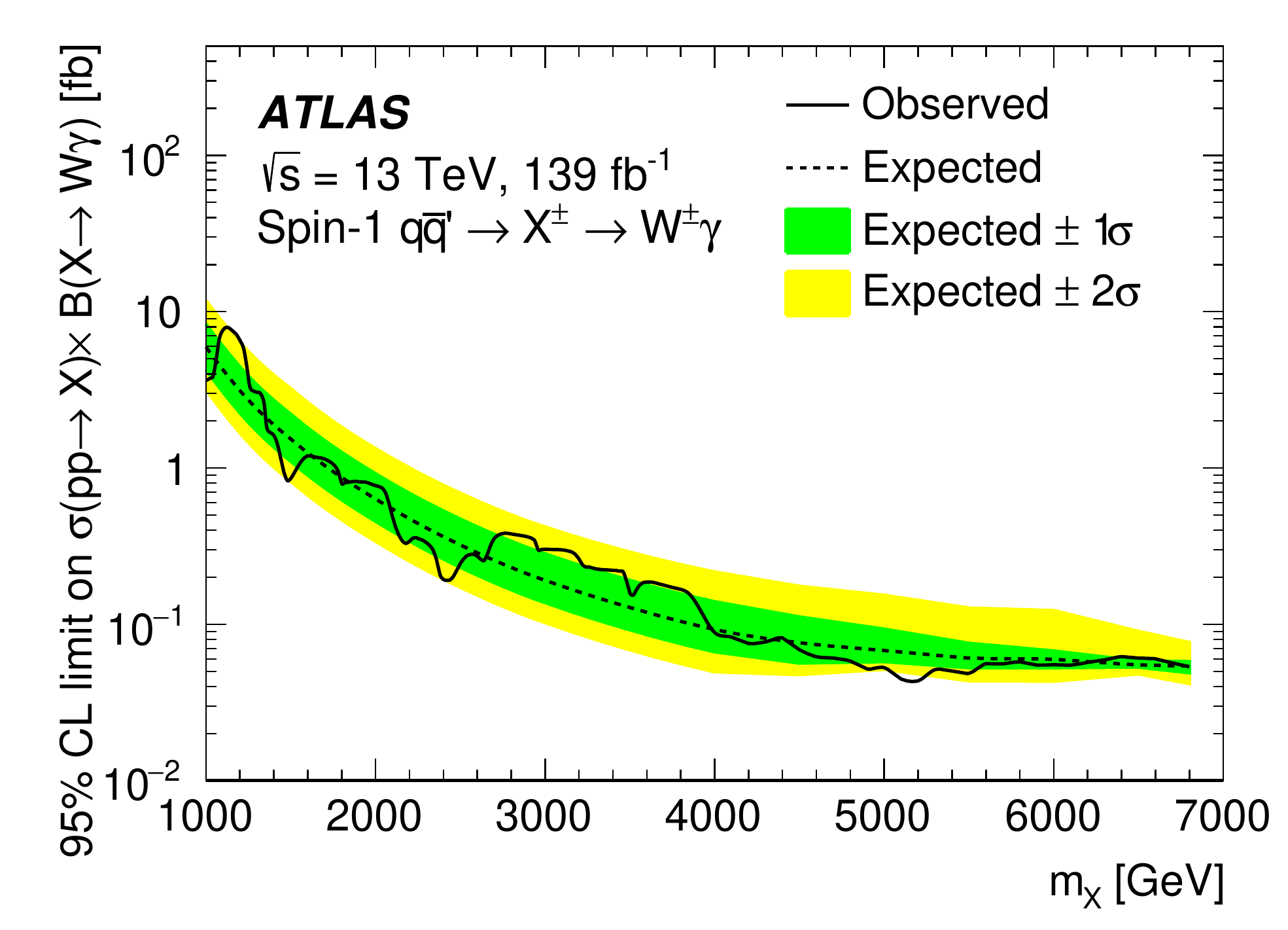}}
\end{center}
\caption{The 95\% CL upper limits on $\sigma(pp\to X)\times B(X\to W/Z\gamma)$ as a function of $m_{X}$ for
(a) \ggXZgJ{0},
(b) \ggXZgJ{2},
(c) \qqXZgJ{2} and
(d) \qqXWgJ{1}.
The observed limits are shown as a solid black line and the expected ones are shown as a dashed line with the 1$\sigma$ (2$\sigma$) uncertainty band presented as the green (yellow) band.
Small discontinuities in $pp\rightarrow X^{0}\rightarrow Z\gamma$ limits are due to dropping the BTAG category from the limit calculation for mass points with $m_{X}>3000\,\GeV$.
Limits for $m_{X}<4000\,\GeV$ are derived with the asymptotic approach, while the ones for higher masses are calculated with the pseudo-experiment sampling method.
}
\label{fig:limits}
\end{figure}

Having found no significant deviation of the data from the SM background predictions, upper limits on signal cross-sections are calculated at a 95\% confidence level for each of the four search channels.
The observed cross-section limits (solid curves) are presented in Figure~\ref{fig:limits}, along with the expected limits (dotted curves) obtained by assuming only SM backgrounds. The limits range between approximately 0.05~fb and 10~fb for $m_{X}$ between 1 and 6.8~\TeV.
The one- and two-standard-deviation bands around the expected limits cover the observed limits almost everywhere, which is consistent with the observation that the data agree well with the background-only expectations.
The largest deviation of the observed cross-section limit above the expectation is 2.5$\sigma$ for \ggXZgJ{0} production from gluon--gluon fusion at $m_{X}=3640$~\GeV.

\FloatBarrier


\section{Conclusion}

Results of searches for high-mass bosons decaying to $W\gamma$ and $Z\gamma$ final states are presented, using 139~fb$^{-1}$ of $\sqrt{s} = 13$~\TeV\ $pp$ collision data collected with the ATLAS detector during the operation of the LHC from 2015 to 2018.
The analysis maximizes the sensitivity of the search by selecting events passing a high-\ET{} photon trigger and identifying jets from the hadronic decays of  highly boosted $W$ and $Z$ bosons.
Distributions of the invariant mass of the photon--jet pairs in the mass range from 1.0 to 6.8~\TeV\ are used to search for $X^\pm \to W^{\pm}\gamma$ and $X^0\to Z \gamma$ signals above a smoothly falling SM background.
No evidence of a new resonance is found, and 95\% confidence-level upper limits on the resonance production cross-section times decay branching fraction are set.
These vary from about 10 to 0.05~fb as the heavy-boson mass increases from 1.0 to 6.8~\TeV. 
Individual studies are carried out for resonances with spin 0, 1, and 2 produced via gluon--gluon fusion and $q\bar{q}$ annihilation, currently providing the most stringent exclusion limits for these processes.
Due to improved analysis techniques, the search sensitivity at high mass has been improved by a factor of two relative to that expected from the increase in integrated luminosity of the analysed data.


\section*{Acknowledgements}


We thank CERN for the very successful operation of the LHC, as well as the
support staff from our institutions without whom ATLAS could not be
operated efficiently.
 
We acknowledge the support of
ANPCyT, Argentina;
YerPhI, Armenia;
ARC, Australia;
BMWFW and FWF, Austria;
ANAS, Azerbaijan;
CNPq and FAPESP, Brazil;
NSERC, NRC and CFI, Canada;
CERN;
ANID, Chile;
CAS, MOST and NSFC, China;
Minciencias, Colombia;
MEYS CR, Czech Republic;
DNRF and DNSRC, Denmark;
IN2P3-CNRS and CEA-DRF/IRFU, France;
SRNSFG, Georgia;
BMBF, HGF and MPG, Germany;
GSRI, Greece;
RGC and Hong Kong SAR, China;
ISF and Benoziyo Center, Israel;
INFN, Italy;
MEXT and JSPS, Japan;
CNRST, Morocco;
NWO, Netherlands;
RCN, Norway;
MEiN, Poland;
FCT, Portugal;
MNE/IFA, Romania;
MESTD, Serbia;
MSSR, Slovakia;
ARRS and MIZ\v{S}, Slovenia;
DSI/NRF, South Africa;
MICINN, Spain;
SRC and Wallenberg Foundation, Sweden;
SERI, SNSF and Cantons of Bern and Geneva, Switzerland;
MOST, Taiwan;
TENMAK, T\"urkiye;
STFC, United Kingdom;
DOE and NSF, United States of America.
In addition, individual groups and members have received support from
BCKDF, CANARIE, Compute Canada and CRC, Canada;
PRIMUS 21/SCI/017 and UNCE SCI/013, Czech Republic;
COST, ERC, ERDF, Horizon 2020 and Marie Sk{\l}odowska-Curie Actions, European Union;
Investissements d'Avenir Labex, Investissements d'Avenir Idex and ANR, France;
DFG and AvH Foundation, Germany;
Herakleitos, Thales and Aristeia programmes co-financed by EU-ESF and the Greek NSRF, Greece;
BSF-NSF and MINERVA, Israel;
Norwegian Financial Mechanism 2014-2021, Norway;
NCN and NAWA, Poland;
La Caixa Banking Foundation, CERCA Programme Generalitat de Catalunya and PROMETEO and GenT Programmes Generalitat Valenciana, Spain;
G\"{o}ran Gustafssons Stiftelse, Sweden;
The Royal Society and Leverhulme Trust, United Kingdom.
 
The crucial computing support from all WLCG partners is acknowledged gratefully, in particular from CERN, the ATLAS Tier-1 facilities at TRIUMF (Canada), NDGF (Denmark, Norway, Sweden), CC-IN2P3 (France), KIT/GridKA (Germany), INFN-CNAF (Italy), NL-T1 (Netherlands), PIC (Spain), ASGC (Taiwan), RAL (UK) and BNL (USA), the Tier-2 facilities worldwide and large non-WLCG resource providers. Major contributors of computing resources are listed in Ref.~\cite{ATL-SOFT-PUB-2021-003}.


\printbibliography
 
\clearpage
 
\begin{flushleft}
\hypersetup{urlcolor=black}
{\Large The ATLAS Collaboration}

\bigskip

\AtlasOrcid[0000-0002-6665-4934]{G.~Aad}$^\textrm{\scriptsize 99}$,
\AtlasOrcid[0000-0002-5888-2734]{B.~Abbott}$^\textrm{\scriptsize 117}$,
\AtlasOrcid[0000-0002-7248-3203]{D.C.~Abbott}$^\textrm{\scriptsize 100}$,
\AtlasOrcid[0000-0002-2788-3822]{A.~Abed~Abud}$^\textrm{\scriptsize 34}$,
\AtlasOrcid[0000-0002-1002-1652]{K.~Abeling}$^\textrm{\scriptsize 53}$,
\AtlasOrcid[0000-0002-2987-4006]{D.K.~Abhayasinghe}$^\textrm{\scriptsize 92}$,
\AtlasOrcid[0000-0002-8496-9294]{S.H.~Abidi}$^\textrm{\scriptsize 27}$,
\AtlasOrcid[0000-0002-9987-2292]{A.~Aboulhorma}$^\textrm{\scriptsize 33e}$,
\AtlasOrcid[0000-0001-5329-6640]{H.~Abramowicz}$^\textrm{\scriptsize 148}$,
\AtlasOrcid[0000-0002-1599-2896]{H.~Abreu}$^\textrm{\scriptsize 147}$
% ...
\bigskip
\\

% ...
$^{27}$Physics Department, Brookhaven National Laboratory, Upton NY; United States of America.\\
% ...
$^{33}$$^{(a)}$Facult\'e des Sciences Ain Chock, R\'eseau Universitaire de Physique des Hautes Energies - Universit\'e Hassan II, Casablanca;$^{(b)}$Facult\'{e} des Sciences, Universit\'{e} Ibn-Tofail, K\'{e}nitra;$^{(c)}$Facult\'e des Sciences Semlalia, Universit\'e Cadi Ayyad, LPHEA-Marrakech;$^{(d)}$LPMR, Facult\'e des Sciences, Universit\'e Mohamed Premier, Oujda;$^{(e)}$Facult\'e des sciences, Universit\'e Mohammed V, Rabat;$^{(f)}$Institute of Applied Physics, Mohammed VI Polytechnic University, Ben Guerir; Morocco.\\
$^{34}$CERN, Geneva; Switzerland.\\
% ...
$^{53}$II. Physikalisches Institut, Georg-August-Universit\"{a}t G\"ottingen, G\"ottingen; Germany.\\
% ...
$^{92}$Department of Physics, Royal Holloway University of London, Egham; United Kingdom.\\
% ...
$^{99}$CPPM, Aix-Marseille Universit\'e, CNRS/IN2P3, Marseille; France.\\
$^{100}$Department of Physics, University of Massachusetts, Amherst MA; United States of America.\\
% ...
$^{117}$Homer L. Dodge Department of Physics and Astronomy, University of Oklahoma, Norman OK; United States of America.\\
% ...
$^{147}$Department of Physics, Technion, Israel Institute of Technology, Haifa; Israel.\\
$^{148}$Raymond and Beverly Sackler School of Physics and Astronomy, Tel Aviv University, Tel Aviv; Israel.\\
% ...
$^{a}$ Also Affiliated with an institute covered by a cooperation agreement with CERN.\\
$^{b}$ Also at Borough of Manhattan Community College, City University of New York, New York NY; United States of America.\\
$^{c}$ Also at Bruno Kessler Foundation, Trento; Italy.\\
$^{d}$ Also at Center for High Energy Physics, Peking University; China.\\
$^{e}$ Also at Centro Studi e Ricerche Enrico Fermi; Italy.\\
$^{f}$ Also at CERN, Geneva; Switzerland.\\
% ...

\end{flushleft}


\clearpage

\end{document}